\documentclass[10pt,journal,compsoc]{IEEEtran}
\ifCLASSOPTIONcompsoc
\usepackage[nocompress]{cite}
\else
\usepackage{cite}
\fi

\usepackage{tabularx}
\usepackage{bm}
\usepackage{calc}

\usepackage{xcolor,colortbl}

\usepackage{array}
\usepackage{rotating}

\usepackage[cmex10]{amsmath}
\usepackage{amssymb}

\DeclareMathOperator*{\softmax}{softmax}

\usepackage{makecell}
\usepackage{subcaption}
\usepackage{textcomp}

\usepackage{multirow}

\usepackage{multicol}

\usepackage{soul}

\usepackage[normalem]{ulem}

\usepackage{algorithm}
\usepackage[noend]{algpseudocode}

\usepackage{mdframed}

\usepackage{url}

\begin{document}
	%
	\title{XSleepNet: Multi-View Sequential Model for Automatic Sleep Staging}
	%
	%
	%
	%
	
	\author{Huy~Phan,
		Oliver~Y.~Ch\'en,
		Minh~C.~Tran,
		Philipp~Koch,
		\\ Alfred~Mertins,
		and~Maarten~De~Vos
		\IEEEcompsocitemizethanks{\IEEEcompsocthanksitem H. Phan is with the School of Electronic Engineering, Queen Mary University of London, London,
			UK, E1 4NS.\protect\\
			E-mail: \texttt{h.phan@qmul.ac.uk}
			\IEEEcompsocthanksitem O. Y. Ch\'en is with the Department of Engineering Science, University of Oxford, Oxford, UK, OX1 3PJ.\protect
			\IEEEcompsocthanksitem M. C. Tran is with Nuffield Division of Anaesthetics, University of Oxford, Oxford, UK, OX3 9DU.\protect
			\IEEEcompsocthanksitem P. Koch and A. Mertins are with the Institute for Signal Processing, University of L\"ubeck, L\"ubeck, Germany, 23562.\protect
			\IEEEcompsocthanksitem M. De Vos is with the Department of Engineering and with the Department of Development and Regeneration, KU Leuven, Leuven, Belgium, 3001.}
	}
	
	%
	%
	
	\markboth{This Article Has Been Published in IEEE Transactions on Pattern Analysis and Machine Intelligence}%
	{This Article Has Been Published in IEEE Transactions on Pattern Analysis and Machine Intelligence}

\IEEEtitleabstractindextext{%
	\begin{abstract}
		Automating sleep staging is vital to scale up sleep assessment and diagnosis to serve millions experiencing sleep deprivation and disorders and enable longitudinal sleep monitoring in home environments. Learning from raw polysomnography signals and their derived time-frequency image representations has been prevalent. However, learning from multi-view inputs (e.g., both the raw signals and the time-frequency images) for sleep staging is difficult and not well understood. This work proposes a sequence-to-sequence sleep staging model, XSleepNet, that is capable of learning a joint representation from both raw signals and time-frequency images. Since different views may generalize or overfit at different rates, the proposed network is trained such that the learning pace on each view is adapted based on their generalization/overfitting behavior. In simple terms, the learning on a particular view is speeded up when it is generalizing well and slowed down when it is overfitting. View-specific generalization/overfitting measures are computed on-the-fly during the training course and used to derive weights to blend the gradients from different views. As a result, the network is able to retain the representation power of different views in the joint features which represent the underlying distribution better than those learned by each individual view alone. Furthermore, the XSleepNet architecture is principally designed to gain robustness to the amount of training data and to increase the complementarity between the input views. Experimental results on five databases of different sizes show that XSleepNet consistently outperforms the single-view baselines and the multi-view baseline with a simple fusion strategy. Finally, XSleepNet also outperforms prior sleep staging methods and improves previous state-of-the-art results on the experimental databases.
	\end{abstract}
	
	\begin{IEEEkeywords}
		Automatic sleep staging, deep neural network, multi-view learning, gradient blending, sequence-to-sequence, end-to-end.
\end{IEEEkeywords}}

\maketitle

\IEEEdisplaynontitleabstractindextext

%
\IEEEpeerreviewmaketitle

\IEEEraisesectionheading{\section{Introduction}\label{sec:introduction}}

\footnotetext[1]{Source code will be available at \url{http://github.com/pquochuy/xsleepnet}.}
Anyone who has experienced a sleepless night would acknowledge the importance of sleep in maintaining one's mental and physical health \cite{Maquet2001,Krieger2017}. Unfortunately, sleep deprivation and disorders are prevalent, affecting millions of people worldwide and imposing serious public health issues \cite{Chattu2019}. For example, 50 to 70 millions of Americans suffer from a chronic sleep or wakefulness disorder, such as insomnia, narcolepsy, restless legs syndrome, and sleep apnea \cite{InstMed2006}. Additionally, medical errors due to sleep deprivation have caused 100 thousand deaths in US hospitals. Consequently, there is an increasing demand for accurate sleep assessment, diagnosis, and longitudinal monitoring in home environments \cite{Mikkelsen2019, Mikkelsen2019b}. In order to address these challenges, it is imperative to employ automated sleep scoring since labor-intensive and time-consuming manual scoring becomes difficult to handle large-scale sleep data. Consider a task to score an overnight polysomnography (PSG) recording. It takes a sleep expert about two hours to complete the task manually \cite{Malhotra2013}; in contrast, a machine can complete it in a few seconds.

The sleep research community is witnessing an unprecedented progress in automatic sleep staging. Machine's performance is approaching sleep experts' \cite{Phan2019a, Stephansen2018}. This is due, in part, to the ever-growing annotated sleep databases. Using large-scale data, novel sleep scoring methods can be developed and tested under powerful deep learning paradigms \cite{Stephansen2018, Oreilly2014, Biswal2018a, Biswal2017, Sun2017}. Since the earlier attempts \cite{Laengkvist2012, Tsinalis2016}, deep learning for automatic sleep staging has evolved rapidly in both designing targeted modelling methodologies and building effective network architectures. The standard one-to-one \cite{phan2018c,phan2018d} and many-to-one \cite{Tsinalis2016, Chambon2018} methods are beginning to be replaced with one-to-many (i.e., multitasking) \cite{Phan2019b} and many-to-many (i.e., sequence-to-sequence) frameworks \cite{Phan2019a, Supratak2017} which better represent the sequential nature of sleep data. Concerning network architectures, the vanilla ones, such as Deep Belief Networks (DBNs) \cite{Laengkvist2012}, Auto-encoders \cite{Tsinalis2016b}, Deep Neural Networks (DNNs) \cite{Dong2017}, Convolutional Neural Networks (CNNs) \cite{Tsinalis2016,Chambon2018,phan2018c, Phan2019b, Andreotti2018, Sors2018}, and Recurrent Neural Networks (RNNs) \cite{phan2018d}, are being surpassed by more complex, task-specific architectures, such as DNN+RNN \cite{Dong2017}, CNN+RNN \cite{Supratak2017}, and hierarchical RNN \cite{Phan2019a}.

Existing work on automatic sleep staging can be categorized based on the types of signal input of the network. There are two main categories: the first directly processes 1-dimensional raw signals \cite{Supratak2017,Dong2017,Chambon2018, Tsinalis2016, Mikkelsen2018, Sors2018, Seo2020} and the second ingests 2-dimensional time-frequency images as inputs \cite{Phan2019a,phan2018c,phan2018d,Stephansen2018}. A time-frequency image is usually derived from a raw signal via some transformations, for example, short-time Fourier transform (STFT). It is, in general, considered as a higher-level representation of the raw signal. However, one cannot conclude that the raw input is better than the time-frequency one as the performance of an automatic sleep staging system depends on many other factors, such as the amount of training data, the network architecture, etc. Rather, they should be considered as two different views regarding the same underlying data distribution. Used together, they should complement each other and improve performance of the task at hand than used separately. Indeed, prior works \cite{Biswal2018a, Biswal2017, Sun2017} have attempted to combine both raw signals and time-frequency images in the same network to tackle automatic sleep staging. Such a network is designed to have a subnet dedicated to an input type. The learned features from different network subnets are then combined (for example, via concatenation) to form joint features on which classification are made. 

In general, learning representations that capture information from multiple views should benefit recognition performance \cite{Zhao2017,Chen2017}. Confusingly, combining multiple input types in a deep network often results in a performance drop rather than an improvement, as we will show in our experiments. This observation has not been well understood.
In this work, we will demonstrate that using a simple strategy like concatenation (as in \cite{Biswal2018a, Biswal2017, Sun2017}) to learn from multi-view input is suboptimal. We will also illustrate why a multi-view network often results in worse performance than the best single-view counterpart. To address this issue, we will introduce a sequence-to-sequence network, XSleepNet, that can learn joint features from both raw and time-frequency input effectively. During training, the network oversees overfitting/generalization behavior on the input views and uses this information to adapt their contributions into the joint-feature learning via gradient blending. Simply put, learning on the view that is generalizing well will be encouraged while learning on the view that is overfitting will be discouraged. In addition, we layout the principles that guides the design of XSleepNet to achieve robustness (to the amount of training data) and complementarity (i.e. how the two input views complement one another). To evaluate the efficacy of the model, we conduct experiments on five databases with different sizes and show that XSleepNet outperforms both three strong baselines and existing works on these databases. It is most likely that XSleepNet is able to consolidate the representation power of the input views to produce the joint features which better represent the underlying data distribution, and therefore, results in higher performance than using single views alone or the simple concatenation fusion strategy.

The rest of the article is organized as follows. 
We outline the principles guiding the network design in Section \ref{ssec:design_principles}. We describe the network architecture and its multi-view joint learning mechanism in Section \ref{sec:xsleepnet}. Details about the experiments will be presented in Section \ref{sec:experiments}, followed by a discussion in Section \ref{ssec:discussion}. We conclude the article in Section~\ref{sec:conclusion}.

\section{Design Principles}
\label{ssec:design_principles}

As a multi-view model, XSleepNet is composed of two network streams: one for the raw signal and the other for the time-frequency image. The following design principles aim to introduce robustness (to the amount of training data) and complementarity (i.e. how the two input views complement one another) into the network while maintaining its flexibility to learn from multiple views effectively.

\emph{Principle 1 (Robustness): The raw-data network stream is large while the time-frequency one is compact in terms of model size.} Specifically, the raw-data network has $5.6\times 10^6$ parameters in total, about 35 times more than $1.6\times10^5$ parameters in the time-frequency one. In general, the footprint of a deep network is proportional to its modelling capacity and should be devised depending on the amount of available training data. The rule of thumb is to increase the network capacity when the training data is large, and decrease it otherwise. However, to do so is not trivial, especially for some clinicians, who may not be technology-savvy. With two network streams of varying modelling capacity, XSleepNet is robust in terms of performance regardless of the amount of training data. When the training data is small, the higher-capacity stream may overfit, but the lower-capacity one generalizes well. When the training data is large, the lower-capacity may underfit, but the higher-capacity stream generalizes well. Combining the two streams results in a balanced, and generalizable model. This is possibly owing to the generalization- and overfitting-aware training procedure of the network (see \emph{Principle 3}). 

\emph{Principle 2 (Complementarity): The two network streams have diverging architectures.} Theoretically, for a joint model to be effective, each individual model should be diversified \cite{Tsymbal2005}. Practically, there is empirical evidence suggesting that CNNs with raw signals and RNNs with time-frequency images are complementary to each other on their sleep staging outputs. For example, the former favors N3, and the latter works better for N1 and REM on MASS database \cite{Phan2019a, Phan2019d}. We, therefore, design XSleepNet such that the raw stream is based on a CNN and the time-frequency stream is based on an RNN to extract epoch-wise features. Even when an RNN is required for inter-epoch sequential modelling, different types of RNN cells are used in the two network streams (see more details in the next section) to ensure the diversification. 

\begin{figure} [!t]
	\centering
	\includegraphics[width=0.9\linewidth]{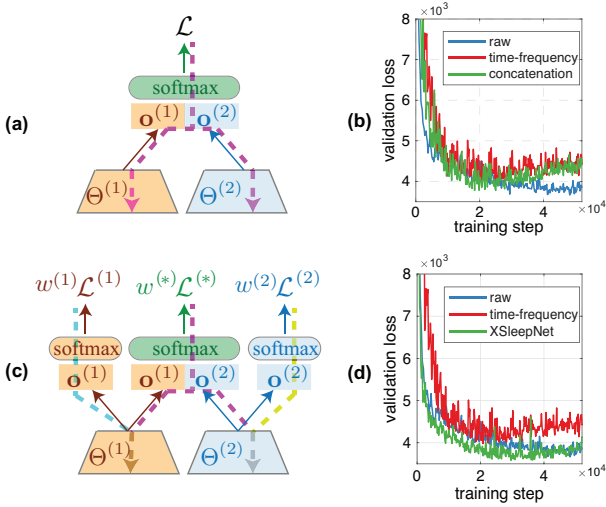}
	\vspace{-0.15cm}
	\caption{Illustration of multi-view learning with simple concatenation and XSleepNet. Joint multi-view learning with simple concatenation (a) and its resulting validation loss in comparison with that of single-view training (b). Joint multi-view learning with XSleepNet (c) and its resulting validation loss in comparison with that of single-view training. In (a) and (c), the dashed lines represent the gradient flows. $\mathbf{\Theta}$, $\mathbf{o}$, and $\mathcal{L}$ denote a network stream corresponding to one input view, a learned feature vector, and a loss value, respectively. The superscripts, i.e. (1) and (2), indicate the input view specifically.  In addition, in (c), $w$ denotes a weight and the superscript (*) indicates the joint network branch.}
	\label{fig:naive_vs_blending}
	\vspace{-0.15cm}
\end{figure}

\emph{Principle 3 (Generalization- and overfitting-aware training): The multi-view network is trained such that learning on the network stream that is generalizing well is accelerated while the overfitting one is discouraged.} This principle is a requirement without which the multi-view network would fail to produce better representation than training either of the single-view network streams separately. In literature, a network with multiple input types typically combines the features learned from the constituent streams (e.g., via concatenation \cite{Biswal2018a, Biswal2017, Sun2017}) to form the joint representation which then serves the classification purpose. This is illustrated in Fig. \ref{fig:naive_vs_blending} (a). As a result, there is not a viable way to regulate the learning pace of the streams individually. This would not be a problem if the network streams generalize and overfit at the same time. However, this is not often the case, as illustrated via their validation losses in Fig. \ref{fig:naive_vs_blending} (b). In turn, the validation loss of the simple combination appears to be averaged out as illustrated in the figure, suggesting worse generalization than the best single-view network stream. 

In order to regulate the learning pace of the network streams, it is necessary to gain access to their gradient flows. In XSleepNet, in addition to the joint classification branch, two additional branches are added. Different from the joint classification branch, these two newly introduced branches operate on the stream-wise features as illustrated in Fig. \ref{fig:naive_vs_blending} (c). By monitoring the generalization/overfitting behavior (see more details in Section \ref{ssec:gradient_bending}) of the classification branches, we are able to weight their gradient flows so that the one that generalizes well is awarded a large weight and the one that overfits is given a small weight. By doing so, we blend the gradients according to the generalization and overfitting behavior of the classification branches and individualize the learning pace of the network streams. Unlike the simple concatenation (see Fig. 1 (b)), with this adaptive gradient blending approach, XSleepNet results in a better joint representation of the underlying data distribution than that of the single-view networks, as evidenced by its validation loss in Fig. \ref{fig:naive_vs_blending} (d).

\section{XSleepNet}
\label{sec:xsleepnet}

\begin{figure*} [!t]
	\centering
	\includegraphics[width=0.8\linewidth]{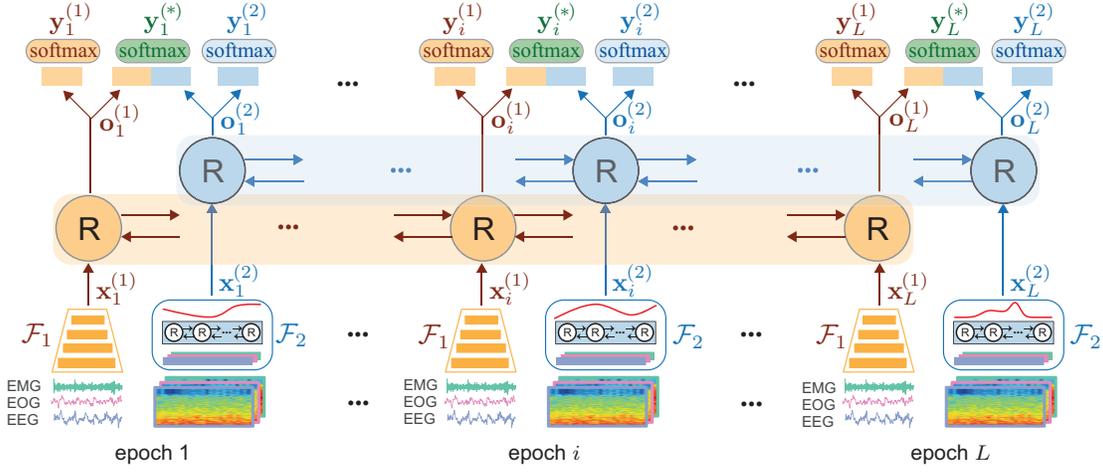}
	\vspace{-0.15cm}
	\caption{The architecture of XSleepNet. The network streams orange and blue correspond to the raw and time-frequency image inputs, respectively.}
	\label{fig:xsleepnet}
	\vspace{-0.15cm}
\end{figure*}

\subsection{Architecture}
\label{ssec:architecture}
Given a training set $\{\mathcal{S}_n\}_{n=1}^N$ of size $N$ where $\mathcal{S}~=~\left(\left\{(\mathbf{S}^{(1)}_1,\mathbf{S}^{(2)}_1), \mathbf{y}_1\right\}, \ldots, \left\{(\mathbf{S}^{(1)}_L,\mathbf{S}^{(2)}_L), \mathbf{y}_L\right\}\right)$ is a sequence of $L$ sleep epochs. $\mathbf{S}^{(1)}_l$ and $\mathbf{S}^{(2)}_l$ represent two different views (i.e., inputs of two different types) and $\mathbf{y}_l \in \{0,1\}^Y$ denotes the one-hot encoding label of the $l$-th epoch, for $1~\le~l~\le~L$. $Y=5$ since we consider the classification of 5 sleep stages in this work. $\mathbf{S}^{(1)}_l \in \mathbb{R}^{3000 \times C}$ is a $C$-channel 30-second raw signal sampled at 100 Hz and $\mathbf{S}^{(2)}_l \in \mathbb{R}^{F \times T \times C}$ is the time-frequency representation, with $F = 129$ frequency bins and $T = 29$ time points (see more details in Section \ref{sec:experiments}). Here, $C$ takes a value of $\{1,2,3\}$ depending on the used channel combination (i.e., EEG, EEG$\cdot$EOG, or EEG$\cdot$EOG$\cdot$EMG).

The architecture of XSleepNet is illustrated in Fig. \ref{fig:xsleepnet}. The network stream in orange handles the raw signal and the one in blue deals with the time-frequency image. Long-term (i.e., inter-epoch) sequential modelling is at the heart of sequence-to-sequence sleep staging models, including XSleepNet. In light of this, we employ bidirectional RNNs for this purpose as in \cite{Phan2019a, Supratak2017}. The raw and time-frequency inputs are therefore encoded into two sequences of output vectors, respectively:
\begin{align}
(\mathbf{o}^{(1)}_1, \ldots,\mathbf{o}^{(1)}_L)&=GRU(\mathcal{F}_1(\mathbf{S}^{(1)}_1), \ldots, \mathcal{F}_1(\mathbf{S}^{(1)}_L)), \label{raw_model} \\
(\mathbf{o}^{(2)}_1, \ldots,\mathbf{o}^{(2)}_L)&=LSTM(\mathcal{F}_2(\mathbf{S}^{(2)}_1), \ldots, \mathcal{F}_2(\mathbf{S}^{(2)}_L)). \label{tf_model}
\end{align}
With respect to the design Principle 2, the bidirectional RNN in (\ref{raw_model}) is realized by Gated Recurrent Unit (GRU) cells \cite{Cho2014} and the one in (\ref{tf_model}) is realized by Long Short-Term Memory (LSTM) cells \cite{Hochreiter1997} coupled with recurrent batch normalization \cite{Cooijmans2016}. In (\ref{raw_model}) and (\ref{tf_model}), $\mathbf{o}^{(1)}_l \!\in\! \mathbb{R}^{2H_1}$ and $\mathbf{o}^{(2)}_l\!\in\!\mathbb{R}^{2H_2}$, for $1 \le l\!\le\!L$, where $H_1$ and $H_2$ are the sizes of the GRU and LSTM cells' hidden state vectors, respectively. $\mathcal{F}_1$ and $\mathcal{F}_2$ denote the subnetworks that play the role of a feature map to transform an input epoch into a feature vector of high-level representation. They are separately tailored for raw and time-frequency inputs with respect to the design Principle 2.

{\bf Feature map $\mathcal{F}_1$:}  In order to transform a raw signal $\mathbf{S}^{(1)}$ into a high-level feature  $\mathbf{x}^{(1)}$, the feature map $\mathcal{F}_1:~\mathbf{S}^{(1)}~\mapsto~\mathbf{x}^{(1)}$ is realized by a \emph{fully} convolutional neural network (FCNN) \cite{Long2015}. Without confusion, the subscript $l$ is omitted here for simplicity. The network consists of 9 strided one-dimensional convolutional layers \cite{Long2015} with a common filter width of 31 and stride length of 2. The convolutional layers are designed to have their numbers of filter increased according to the network's depth, taking values of $16, 16, 32, 32, 64, 64, 128, 128,$ and $256$, to compensate for the gradually smaller induced feature maps. Given the input $\mathbf{S}^{(1)}$ of size $3000 \times C$, the CNN results in feature map sizes of $1500 \times 16$, $750 \times 16$, $325 \times 32$, $163 \times 32$, $82 \times 64$, $41 \times 64$, $21 \times 128$, $11 \times 128$, and $6 \times 256$ after the 9 convolutional layers, respectively. In addition, each convolutional layer is followed by parametric rectified linear units (PReLUs) \cite{He2015}. 
The output of the last convolutional layer is flattened to form the induced epoch-wise feature vector $\mathbf{x}^{(1)} = \mathcal{F}_1\left(\mathbf{S}^{(1)}\right) \in \mathbb{R}^{1536}$. 

{\bf Feature map $\mathcal{F}_2$:}  Different from $\mathcal{F}_1$, the feature map $\mathcal{F}_2:~\mathbf{S}^{(2)}~\mapsto~\mathbf{x}^{(2)}$ relies on the attention-based RNN coupled with learnable \emph{filterbank layers} to map a time-frequency input $\mathbf{S}^{(2)}$ to a high-level feature vector $\mathbf{x}^{(2)}$. Again, the subscript $l$ is omitted for simplicity. First, each channel of $\mathbf{S}^{(2)}$ is preprocessed by a learnable filterbank layer of $D$ filters as in \cite{Phan2019a, phan2018c} so that its spectral dimension is smoothed and reduced from $F$ to $D$ frequency bins, resulting in $\mathbf{S'}^{(2)} \in \mathbb{R}^{T \times D \times C}$. Afterwards, $\mathbf{S'}^{(2)}$ is interpreted as a sequence of $T$ vectors $(\mathbf{s'}_1, \ldots, \mathbf{s'}_T)$, where each $\mathbf{s'}_t \in \mathbb{R}^{DC}$, for $1 \le t \le T$, and encoded by a bidirectional RNN into a sequence of vectors $(\mathbf{z}_1, \ldots, \mathbf{z}_T)$:
\begin{align}
(\mathbf{z}_1, \ldots, \mathbf{z}_T) = LSTM(\mathbf{s'}_1, \ldots, \mathbf{s'}_T).
\end{align}
Here the bidirectional RNN is realized by LSTM cells \cite{Hochreiter1997} coupled with recurrent batch normalization \cite{Cooijmans2016} and $\mathbf{z}_t \in \mathbb{R}^{2H}$ with $H$ equal to the size of the LSTM cells' hidden state vectors. Note that this bidirectional RNN here is for short-term (i.e. intra-epoch) sequential modelling and should not be confused with the one for inter-epoch sequential modelling in (\ref{tf_model}). The induced epoch-wise feature vector $\mathbf{x}^{(2)}$ is eventually obtained via a weighted combination of the vectors $\mathbf{z}_1, \ldots, \mathbf{z}_T$:
\begin{align}
\mathbf{x}^{(2)} = \sum\nolimits_{t=1}^T \alpha_t\mathbf{z}_t. \label{eq:attention_feat}
\end{align}
In (\ref{eq:attention_feat}), $\alpha_1, \ldots, \alpha_T$ are attention weights learned by an attention layer as in \cite{phan2018d,Phan2019a}:
\begin{align}
\alpha_t &= \frac{\exp(\mathbf{a}_t^{\mathsf{T}}\mathbf{a}_e)}{\sum_{t=1}^{T}\exp(\mathbf{a}_t^{\mathsf{T}}\mathbf{a}_e)}, \label{eq:attention_weights} \\ \mathbf{a}_t &= \tanh(\mathbf{W}_a\mathbf{z}_t + \mathbf{b}_a),
\end{align}
where $\mathbf{W}_a\in \mathbb{R}^{A\times2H}$ and $\mathbf{b}_a\in \mathbb{R}^{A}$ are trainable weight matrix and bias vector, respectively. $\mathbf{a}_e\in \mathbb{R}^{A}$ is the trainable epoch-level context vector. $A$ is the so-called attention size.

In light of the design Principle 2, it is worth noting that the ways $\mathcal{F}_1$ and $\mathcal{F}_2$ learn to produce features are distinguishable. On the one hand, we conjecture that, due to its reliance on FCNN, $\mathcal{F}_1$ can capture temporal-equivariant patterns from the raw input. That is, such a feature can occur at any time point in the 30-second duration of the raw signal. For example, micro events such as K-complex and sleep spindle \cite{Iber2007,Hobson1969} frequently appearing in the sleep stage N2 have this characteristic. On the other hand, built upon the bidirectional RNN and the attention mechanism, $\mathcal{F}_2$ can encode the sequential patterns of the spectral columns in its time-frequency image input. This is useful to capture features with a sequential structure, such as the theta wave activity in the sleep stage N1 \cite{Iber2007,Hobson1969}.

Adhering to the design Principle 3, XSleepNet accommodates three softmax branches for classification: the first two are view-specific (i.e., operating on the output vectors of the raw and time-frequency network streams specifically) and the third one for the joint view (i.e., operating on the joint feature vector), as illustrated in Fig. \ref{fig:naive_vs_blending}. Given the output sequence $(\mathbf{o}^{(1)}_1, \ldots,\mathbf{o}^{(1)}_L)$ in (\ref{raw_model}) and $(\mathbf{o}^{(2)}_1, \ldots,\mathbf{o}^{(2)}_L)$ in (\ref{tf_model}), we obtain three sequences of classification labels $(\hat{\mathbf{y}}^{(1)}_1, \ldots,\hat{\mathbf{y}}^{(1)}_L)$, $(\hat{\mathbf{y}}^{(2)}_1, \ldots,\hat{\mathbf{y}}^{(2)}_L)$, and $(\hat{\mathbf{y}}^{(*)}_1, \ldots,\hat{\mathbf{y}}^{(*)}_L)$ via the three classification layers, where
\begin{align}
\hat{\mathbf{y}}^{(1)}_l &= \softmax(\mathbf{o}^{(1)}_l ), \label{classification_raw}\\
\hat{\mathbf{y}}^{(2)}_l &= \softmax(\mathbf{o}^{(2)}_l ), \label{classification_tf}\\
\hat{\mathbf{y}}^{(*)}_l &= \softmax([\mathbf{o}^{(1)}_l \oplus \mathbf{o}^{(2)}_l]), 1 \le l \le L, \label{classification_joint}
\end{align}
and $\oplus$ denotes vector concatenation.

During training, the losses induced by these three classification branches are used to compute the weights for gradient blending (cf. Section \ref{ssec:gradient_bending}). When the trained network is evaluated on the test data, the joint output $(\hat{\mathbf{y}}^{(*)}_1, \ldots,\hat{\mathbf{y}}^{(*)}_L)$ 
are considered as the final outputs. 

\subsection{Adaptive Gradient Blending for Multi-View Training}
\label{ssec:gradient_bending}

Given three classification outputs $(\hat{\mathbf{y}}^{(1)}_1, \ldots,\hat{\mathbf{y}}^{(1)}_L)$, $(\hat{\mathbf{y}}^{(2)}_1, \ldots,\hat{\mathbf{y}}^{(2)}_L)$, and $(\hat{\mathbf{y}}^{(*)}_1, \ldots,\hat{\mathbf{y}}^{(*)}_L)$, the cross-entropy losses induced by the three classification branches are
\begin{align}
\mathcal{L}^{(k)} = -\frac{1}{L}\sum\nolimits_{l=1}^L\mathbf{y}_l\log(\hat{\mathbf{y}}_l^{(k)}),
\end{align}
where $k \in \{1,2,*\}$. In order to balance the generalization/overfitting rate of the classification branches, we weight the losses adaptively to schedule the learning on the two network streams. Similar to \cite{Wang2020}, we computed the loss weights using the ratio of generalization and overfitting measures (see Appendix A for a theoretical justification):
\begin{align}
w^{(k)} = \frac{1}{Z}\frac{G_k}{O_k^2}. \label{weight}
\end{align}
In (\ref{weight}), $Z$ is a normalization factor. The generalization measure $G_k$ is defined as the gained information about the target distribution we learn from training. The overfitting measure $O_k$ is defined as the gap between the gain on the training set and the target distribution. The weighted loss used for training at the training step $n$ is then given by:
\begin{align}
\mathcal{L}(n) = \sum_{k\in\{1,2,*\}}w^{(k)}(n)\mathcal{L}^{(k)}(n). \label{total_loss}
\end{align}

\begin{figure} [!t]
		\centering
		\includegraphics[width=0.9\linewidth]{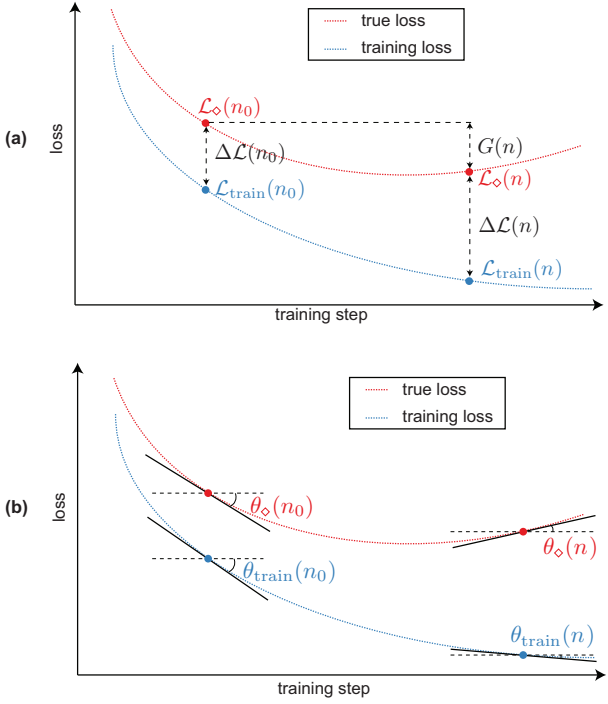}
		\vspace{-0.15cm}
		\caption{Theoretical illustration of the interrelation between $\mathcal{L}_{\text{train}}^{(k)}(n)$,  $\mathcal{L}_{\diamond}^{(k)}(n)$, $\mathcal{L}_{\text{train}}^{(k)}(n_0)$ and $\mathcal{L}_{\diamond}^{(k)}(n_0)$ in XSleepNet1 (a); and the interrelation between $\theta_{\text{train}}^{(k)}(n)$,  $\theta_{\diamond}^{(k)}(n)$, $\theta_{\text{train}}^{(k)}(n_0)$ and $\theta_{\diamond}^{(k)}(n_0)$ in XSleepNet2 (b).}
		\label{fig:xsleepnet1_vs_xsleepnet2}
	\vspace{-0.25cm}
\end{figure}

Using the weighted loss function in (\ref{total_loss}), the network's learning behavior depends on how $G_k$ and $O_k$ in (\ref{weight}) are approximated. We present two approximation approaches for XSleepNet, resulting in two algorithms, namely \emph{XSleepNet1} and \emph{XSleepNet2}.
	
{\bf XSleepNet1}: Inspired by \cite{Wang2020}, we approximate $G_k$ and $O_k$ at the training step $n$ as:
\begin{align}
G_k(n)\! &\approx\! \mathcal{L}_{\diamond}^{(k)}(n_0) - \mathcal{L}_{\diamond}^{(k)}(n), \label{eq:Gx}\\
O_k\!(n)\! &\approx\! \Delta\mathcal{L}(n) - \Delta\mathcal{L}(n_0) \nonumber \\ 
&\approx\! \!\left[\!\mathcal{L}_{\diamond}^{(k)}\!(n) - \mathcal{L}_{\text{train}}^{(k)}\!(n)\!\right] - \left[\!\mathcal{L}_{\diamond}^{(k)}\!(n_0) - \mathcal{L}_{\text{train}}^{(k)}\!(n_0)\!\right],\label{eq:Ox}
\end{align}
where $\mathcal{L}_{\text{train}}^{(k)}(n)$ and $\mathcal{L}_{\diamond}^{(k)}(n)$ denote the training loss and the true loss at the training step $n$, respectively. $\mathcal{L}_{\text{train}}^{(k)}(n_0)$ and $\mathcal{L}_{\diamond}^{(k)}(n_0)$, where $n_0 < n$, denote the training and true loss references, respectively. The interrelation between $\mathcal{L}_{\text{train}}^{(k)}(n)$,  $\mathcal{L}_{\diamond}^{(k)}(n)$, $\mathcal{L}_{\text{train}}^{(k)}(n_0)$ and $\mathcal{L}_{\diamond}^{(k)}(n_0)$ is illustrated in Fig. \ref{fig:xsleepnet1_vs_xsleepnet2} (a). As in \cite{Wang2020}, we set $n_0 = 0$ (i.e., right after the network has been initialized with random weights).

{\bf XSleepNet2}: There are at least two potential drawbacks in the approximation approach used in XSleepNet1. First, in theory, the (training and validation) loss curves are smooth (as illustrated in Fig. \ref{fig:xsleepnet1_vs_xsleepnet2}), yet they are typically noisy in practice (cf. Fig. \ref{fig:naive_vs_blending}) due to minibatch training. As a result, using the spontaneous loss values $\mathcal{L}_{\text{train}}^{(k)}(n)$ and $\mathcal{L}_{\diamond}^{(k)}(n)$ in the approximation leads to considerably varying loss weights which result in unstable learning behavior and suboptimal performance, particularly when the data size is small. Second, the references $\mathcal{L}_{\text{train}}^{(k)}(0)$ and $\mathcal{L}_{\diamond}^{(k)}(0)$ may vary significantly with different random initializations. This variation further adds up to the approximation suboptimality.

\begin{algorithm}[t!]
		\footnotesize
		\caption{Procedure for computing the loss weight using XSleepNet2's approximation.}
		\label{alg:xsleepnet2}
		\begin{algorithmic}[1]
			
			\Procedure{XSleepNet2\_Weight}{$L_{\text{train}}, L_\diamond, W, n_0$}       
			\State {\bf input:} $L_{\text{train}}[1\ldots n]$: array of training loss values \\
			{~~~~~~~~~~~~~~~~}$L_\diamond[1\ldots n]$: array true loss values \\
			{~~~~~~~~~~~~~~~~}$W$: window size for line fitting
			\State {\bf output:} $w(n)$: weight at the training time $n$ \\ 
			{~~~~~~~~~~~~~~~~~~} $n_0$: updated reference time step
			\If{$n < W$}
			\Return
			\EndIf
			\State $\tan{\theta}_{\text{train}}(n) \leftarrow \text{LineFit}(L_{\text{train}}[(n-W) \ldots n])$ 
			\State $\tan{\theta}_{\diamond}(n) \leftarrow \text{LineFit}(L_{\diamond}[(n-W) \ldots n])$ 
			\State $G(n) = \tan\theta_{\diamond}\!(n)\!-\!\tan\theta_{\diamond}\!(n_0)$ 
			\State $O(n) = \left[\tan\theta_{\diamond}\!(n)\!-\!\tan\theta_{\text{train}}\!(n)\right]\!-\!\left[\tan\theta_{\diamond}\!(n_0)\!-\!\tan\theta_{\text{train}}\!(n_0)\right]$ 
			\State $w(n) = \frac{1}{Z}\frac{G(n)}{O^2(n)}$  
			\If{$\tan{\theta}_{\diamond}(n_0) > \tan{\theta}_{\diamond}(n)$}
			\State{$n_0 = n$} \Comment{Time step w.r.t. the true loss's smallest tangent}
			\EndIf
			\EndProcedure
		\end{algorithmic}
\end{algorithm}

In XSleepNet2, we introduce a novel approximation approach relying on tangents of the losses, which are in turn robustly approximated via line fitting, to overcome these drawbacks. This approach is based on two observations about the loss curves. First, although noisy, a loss curve's overall trend is smooth (i.e., changing slowly). Furthermore, the loss trend at a training step $n$ can be well represented by the direction of its tangent line at $n$. Second, the directions of the loss tangent lines at $n$ are indicative of the network's generalization/overfitting behavior. Let $\theta_{\text{train}}^{(k)}\!(n)$, and $\theta_{\diamond}^{(k)}\!(n)$, where $-90^\circ\!\le\!\theta_{\text{train}}^{(k)}\!(n)\!\le\!0^\circ$ and $-90^\circ\!\le\!\theta_{\diamond}^{(k)}\!(n)\!\le\!90^\circ$, denote the angles made by the tangent lines of the training and true loss curves with the horizontal axis, respectively, as illustrated in Fig \ref{fig:xsleepnet1_vs_xsleepnet2} (b). The network is generalizing when $-90^\circ\!\le\!\theta_{\diamond}^{(k)}\!(n)\!<\!0^\circ$ (i.e., negative tangent) and overfitting when $0^\circ\!<\!\theta_{\diamond}^{(k)}\!(n)\!\le\!90^\circ$ (i.e., positive tangent). In this light, $G_k$ and $O_k$ are approximated as: 
\begin{align}
G_k(n) &\approx \tan\theta_{\diamond}^{(k)}(n) - \tan\theta_{\diamond}^{(k)}(n_0), \label{eq:Gx_tangent}\\
O_k(n) &\approx \left[\tan\theta_{\diamond}^{(k)}(n) - \tan\theta_{\text{train}}^{(k)}(n)\right] \nonumber \\
&{~~~~} - \left[\tan\theta_{\diamond}^{(k)}(n_0) - \tan\theta_{\text{train}}^{(k)}(n_0)\right], \label{eq:Ox_tangent}
\end{align}where $\tan\theta_{\text{train}}^{(k)}(n)$ and $\tan\theta_{\diamond}^{(k)}(n)$ are the training and true loss tangents at the training step $n$ and $\tan\theta_{\text{train}}^{(k)}(n_0)$ and $\tan\theta_{\diamond}^{(k)}(n_0)$ are their references, respectively. Since a loss tangent is the ratio of the loss change to the increase of the training step (i.e., the slope of the tangent line), the approximations in (\ref{eq:Gx_tangent}) and (\ref{eq:Ox_tangent}) are second-order whereas those in (\ref{eq:Gx}) and (\ref{eq:Ox}) (i.e., in XSleepNet1) are first-order.
	
The procedure for computing the adaptive loss weight $w^{(k)}$ of the network branch $k$ using XSleepNet2's approximation is described in Algorithm \ref{alg:xsleepnet2}. We approximate a loss tangent (i.e., $\tan\theta_{\text{train}}^{(k)}(n)$ and $\tan\theta_{\diamond}^{(k)}(n)$) at the time step $n$ by the slope of the lines fitted to the loss curve segment of length $W$, starting from the time step $n-W$ and proceeding up to $n$. Because of this, the network training starts with a warm-up period of length $W$ during which the weights of all the network branches are set to be equal. In addition, on each network branch, the time step with respect to the smallest true loss tangent (i.e., the steepest slope of the fitting line) is used as the reference time step $n_0$.

In practice, in both XSleepNet1 and XSleepNet2 the true loss $\mathcal{L}_{\diamond}^{(k)}(n)$ is unknown. We, therefore, approximate it by the loss evaluated on a held-out validation set, $\mathcal{L}_{\text{valid}}^{(k)}(n)$. To reduce the computational cost of obtaining the training loss $\mathcal{L}_{\text{train}}^{(k)}(n)$ on the entire training set, only a subset randomly sampled from the training set is used for this purpose. Note that, although the second-order approximations in (\ref{eq:Gx_tangent}) and (\ref{eq:Ox_tangent}) require more computation than the first-order ones in (\ref{eq:Gx}) and (\ref{eq:Ox}), the extra cost is negligible compared to the networks' computation. Hence, there is almost no difference between the networks in terms of computational cost.

\vspace{-0.1cm}
\section{Experiments}
\label{sec:experiments}

\setlength\tabcolsep{2.25pt}
\begin{table*}[!t]
	\caption{Summary of the employed databases.}
	\vspace{-0.35cm}
	\scriptsize
	\begin{center}
		\begin{tabular}{|>{\arraybackslash}m{0.6in}|>{\centering\arraybackslash}m{0.35in}|>{\centering\arraybackslash}m{0.7in}|>{\centering\arraybackslash}m{0.7in}|>{\centering\arraybackslash}m{0.85in}|>{\centering\arraybackslash}m{0.7in}|>{\centering\arraybackslash}m{0.9in}|>{\centering\arraybackslash}m{0.7in}|>{\centering\arraybackslash}m{0.75in}|>{\centering\arraybackslash}m{0in} @{}m{0pt}@{}}
			\cline{1-9}
			Database & Size & EEG channel & EOG channel & EMG channel & Scoring method & Experimental setup & Held-out validation set & Random training subset &\parbox{0pt}{\rule{0pt}{1ex+\baselineskip}} \\ [0ex]  	
			
			\cline{1-9}
			SleepEDF-20 & 20 & Fpz-Cz & ROC-LOC & $-$ & R\&K & 20-fold CV & 4 subjects & 4 subjects &\parbox{0pt}{\rule{0pt}{0.25ex+\baselineskip}} \\ [0ex]  	
			SleepEDF-78 & 79 & Fpz-Cz  & ROC-LOC & $-$ & R\&K & 10-fold CV & 7 subjects & 7 subjects &\parbox{0pt}{\rule{0pt}{0.25ex+\baselineskip}} \\ [0ex]  	
			MASS & 200 & C4-A1/C3-A2 & ROC-LOC & CHIN1-CHIN2 & AASM/R\&K & 20-fold CV & 10 subjects & 10 subjects &\parbox{0pt}{\rule{0pt}{0.25ex+\baselineskip}} \\ [0ex]  	
			Physio2018 & 994 & C3-A2 & E1-M2 & CHIN1-CHIN2 & AASM & 5-fold CV & 50 subjects & 50 subjects &\parbox{0pt}{\rule{0pt}{0.25ex+\baselineskip}} \\ [0ex]  	
			SHHS &  5,791 & C4-A1 & ROC-LOC & Submental EMG & R\&K & train/test: 0.7/0.3 & 100 subjects & 100 subjects &\parbox{0pt}{\rule{0pt}{0.25ex+\baselineskip}} \\ [0ex]  	
			\cline{1-9}
		\end{tabular}
	\end{center}
	\label{tab:databases}
	\vspace{-0.4cm}
\end{table*}

\subsection{Databases}

We employed five publicly available databases in this study: SleepEDF-20, SleepEDF-78, Montreal Archive of Sleep Studies (MASS), Physonet2018, and Sleep Heart Health Study (SHHS). A summary of the databases is shown in Table \ref{tab:databases}.

{\bf SleepEDF-20:} This is the Sleep Cassette (SC) subset of the Sleep-EDF Expanded dataset \cite{Kemp2000, Goldberger2000} (version 2013), consisting of 20 subjects (10 males and 10 females) aged 25-34. Two consecutive day-night PSG recordings were collected for each subject, except for subject 13 who had one night's data lost due to device failure. Each 30-second PSG epoch was manually labelled into one of eight categories \{W, N1, N2, N3, N4, REM, MOVEMENT, UNKNOWN\} by sleep experts according to the R\&K standard \cite{Hobson1969}. Similar to previous works \cite{Tsinalis2016, Tsinalis2016b, Supratak2017, Phan2019b, phan2018c, phan2018d}, N3 and N4 stages were considered as N3 collectively and MOVEMENT and UNKNOWN categories were excluded. We adopted the Fpz-Cz EEG and ROC-LOC EOG (i.e., the EOG horizontal) channels in this study. However, we did not experiment with EMG as full EMG recordings are not available.

{\bf SleepEDF-78:} This database is the 2018 version of the Sleep-EDF Expanded dataset \cite{Kemp2000, Goldberger2000}, consisting of 78 healthy Caucasian subjects aged 25-101. Similar to SleepEDF-20, two consecutive day-night PSG recordings were collected for each subject, except subjects 13, 36, and 52 whose one recording was lost due to device failure. Manual scoring was done by sleep experts according to the R\&K standard \cite{Hobson1969} and each 30-second PSG epoch was labeled as one of eight categories \{W, N1, N2, N3, N4, REM, MOVEMENT, UNKNOWN\}. N3 and N4 stages were merged into N3 stage. MOVEMENT and UNKNOWN epochs were excluded. We used the Fpz-Cz EEG and ROC-LOC EOG in this study and no experiments were carried out with EMG due to its unavailability.

{\bf MASS:} This database was pooled from different hospital-based sleep laboratories, consisting of whole-night recordings from 200 subjects (97 males and 103 females) aged 18-76. Manual annotation was accomplished by sleep experts according to the AASM standard \cite{Iber2007} (SS1 and SS3 subsets) or the R\&K standard \cite{Hobson1969}  (SS2, SS4, and SS5 subsets). As in \cite{Phan2019b,Phan2019a}, we converted R\&K annotations into five sleep stages \{W, N1, N2, N3, REM\} according to the AASM standard. Epochs with a length of 20 seconds were expanded to 30-second ones by including 5-second segments before and after them. We used C4-A1 EEG, ROC-LOC EOG, and CHIN1-CHIN2 EMG in our experiments. 

{\bf Physio2018:} This database was contributed by Massachusetts General Hospital’s Computational Clinical Neurophysiology Laboratory. It was used in the 2018 Physionet challenge \cite{Ghassemi2018,Goldberger2000} to detect arousal during sleep. We employed the training set (annotation of the evaluation set was not publicly available) consisting of 944 subjects aged 18-90 in the experiments. Manual scoring was done by sleep experts according to the AASM guideline \cite{Iber2007}. C3-A2 EEG, E1-M2 EOG, and CHIN1-CHIN2 EMG were used. 

{\bf SHHS:} The SHHS database \cite{Zhang2018, Quan1997} was collected from multiple centers to study the effect of sleep-disordered breathing on cardiovascular diseases. It has two rounds of PSG records, namely Visit 1 (SHHS-1) and Visit 2 (SHHS-2). The former, consisting of 5,791 subjects aged 39-90, was employed in this work. Manual scoring was completed using the R\&K guideline \cite{Hobson1969}. Similar to other databases annotated with the R\&K rule, N3 and N4 stages were merged into N3 stage and MOVEMENT and UNKNOWN epochs were discarded. We adopted C4-A1 EEG, ROC-LOC EOG, and bipolar submental EMG in the experiments.

These databases were adopted in this work to show the robustness of XSleepNet in dealing with datasets of different sizes, which is one of the guiding principles of the network design (see more details in Section \ref{ssec:design_principles}). For each database, we experimented with single-channel EEG, 2-channel EEG$\cdot$EOG, and 3-channel  EEG$\cdot$EOG$\cdot$EMG combinations as inputs where possible. Particularly, only single-channel and 2-channel experiments were conducted on SleepEDF-20 and SleepEDF-78 since they do not have a full EMG channel available in the PSG recordings. All the signals were resampled to 100 Hz. 

We conducted experiments following the data splits as summarized in Table \ref{tab:databases}. These data splits have been commonly used in literature, enabling a direct performance comparison between our system and prior works. As noted in the table, a number of subjects were held out for validation. Particularly, for the SHHS database, 
we did not leave out a validation set as large as 20\% of the subjects as in Sors \emph{et al.} \cite{Sors2018}; rather, we left out 100 subjects as we empirically found 100 subjects were sufficient for validation purpose.
Furthermore, a random subset was also drawn from the training subjects to compute the loss weights as described in Section \ref{ssec:gradient_bending}. Of note, different from the validation set, which is not involved in network training, this training subset contributes to the training process as usual. 

\vspace{-0.1cm}
\subsection{Parameters}
\label{ssec:parameters}

To extract the time-frequency input mentioned in Section \ref{ssec:parameters}, a signal (i.e., EEG or EOG or EMG) of a 30-second PSG epoch sampled at 100 Hz was divided into two-second windows with 50\% overlap, multiplied with a Hamming window, and transformed to the frequency domain using a 256-point Fast Fourier Transform (FFT). The amplitude spectrum was then log-transformed. The time-frequency images extracted from a database were normalized to zero mean and unit standard deviation before training and testing.

The network implementation was based on \emph{Tensorflow} framework \cite{Abadi2016}. The input sequence length was set to $L\!=\!20$ which was proven to be a reasonable value for sequence-to-sequence models \cite{Phan2019a}. The feature map $\mathcal{F}_2$ was designed with $D\!=\!32$ filters in each filterbank layer, 64 units in its LSTM cells' hidden state vectors, and the attention size $A\!=\!64$ (see Appendix E for an ablation study on the epoch-level attention weights in (\ref{eq:attention_weights})). The LSTM and GRU cells responsible for inter-epoch sequential modelling had 256 and 64 units in their hidden state vectors, respectively. Throughout the network, dropout \cite{Srivastava2014} was applied during training with a dropout rate of $0.5$ and $0.25$ used for convolutional layers and recurrent layers of the network, respectively. 

The network was trained for 10 epochs using Adam optimizer \cite{Kingma2015} with a learning rate of $10^{-4}$, $\beta_1\!=\!0.9$,  $\beta_2\!=\!0.999$, and $\epsilon\!=\!10^{-7}$. The minibatch size of 32 was used for training. The model was validated on the validation set every 100 training steps and the loss weights were also updated at the same time. The parameter $W$ (i.e., the window for line fitting and the warm-up period) in Algorithm \ref{alg:xsleepnet2} was set to 20 evaluation steps. We also smoothed the loss curves with $W$-point moving average before line fitting. For the larger databases Physio2018 and SHHS, the training process was early stopped after 200 validation times if no accuracy improvement was recorded on the validation set. Early stopping was not exercised for other databases.

\vspace{-0.1cm}
\subsection{Baseline systems}
\label{ssec:baselines}

To assess the efficacy of XSleepNet1 and XSleepNet2, in addition to relevant prior works, we developed three baseline systems for performance comparison. (1) The first two baselines, \emph{FCNN+RNN} and \emph{ARNN+RNN}, are equivalent to the raw and time-frequency network streams shown in Fig. \ref{fig:xsleepnet}, respectively, but they were trained separately. Note that   the \emph{ARNN+RNN} is similar to SeqSleepNet presented in \cite{Phan2019a}. The third baseline, denoted as \emph{Naive Fusion}, combines the multi-view features in a concatenate fashion (i.e., without adaptive gradient blending) and shares a similar network architecture to that of XSleepNet.

The networks were evaluated using five metrics, including accuracy, macro F1-score (MF1) \cite{Yang1999}, Cohen's kappa \cite{McHugh2012}, average sensitivity, and average specificity. The outputs of a system across the cross-validation folds were pooled altogether to compute its performance metrics. Statistics of the performance metrics, including mean and standard deviation, were also reported in Appendix B.

\begin{figure*} [!t]
	\centering
	\includegraphics[width=0.8\linewidth]{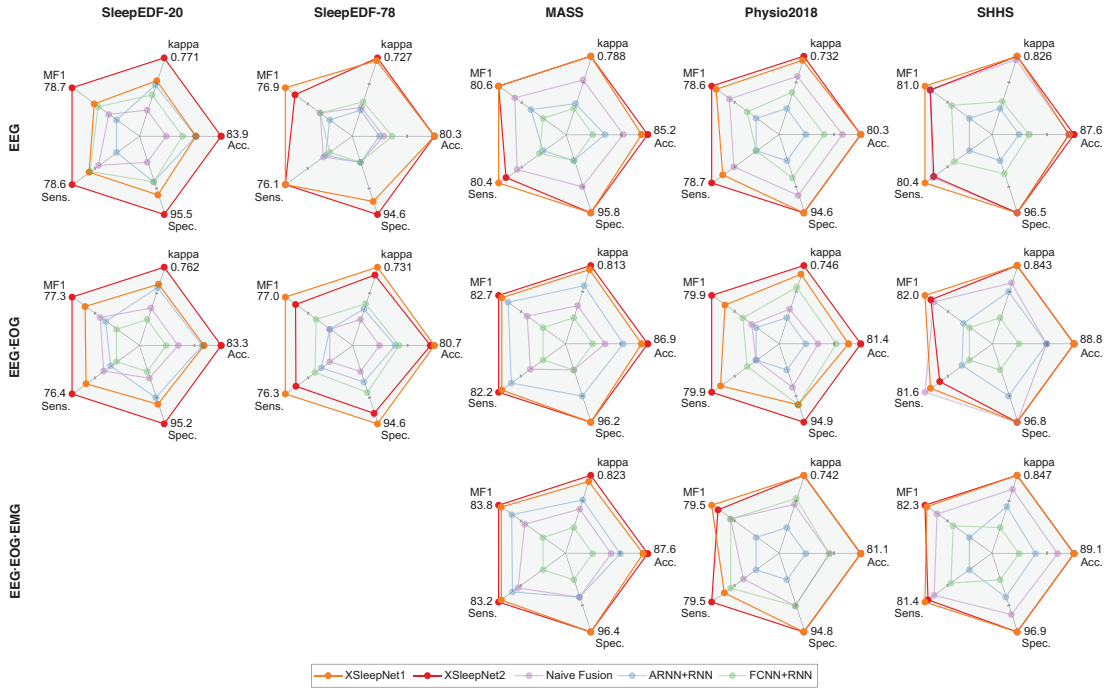}
	\vspace{-0.15cm}
	\caption{An overview on the overall performance obtained by XSleepNet1, XSleepNet2, and the developed baselines.}
	\label{fig:overall_performance}
	\vspace{-0.25cm}
\end{figure*}

\setlength\tabcolsep{2.25pt}
\begin{table*}[!t]
	\caption{Performance comparison between XSleepNet1, XSleepNet2, the baselines, and previous works on the experimental databases. Of note, the \emph{ARNN+RNN} baseline is equivalent to SeqSleepNet presented in \cite{Phan2019a}. In addition, the results indicated by $^\dagger$ are not directly comparable since they either relied on transfer learning with a pretrained model or used a different subset of the corresponding databases. We include them here for the sake of completeness. We mark in bold where XSleepNets' performances are better or equal to those of all the baselines and compatible prior works.}
\vspace{-0.25cm}
	\scriptsize
	\begin{center}
		\begin{tabular}{|>{\centering\arraybackslash}m{0.55in}|>{\arraybackslash}m{1.4in}||>{\centering\arraybackslash}m{0.25in}|>{\centering\arraybackslash}m{0.3in}|>{\centering\arraybackslash}m{0.25in}|>{\centering\arraybackslash}m{0.25in}|>{\centering\arraybackslash}m{0.25in}||>{\centering\arraybackslash}m{0.25in}|>{\centering\arraybackslash}m{0.3in}|>{\centering\arraybackslash}m{0.25in}|>{\centering\arraybackslash}m{0.25in}|>{\centering\arraybackslash}m{0.25in}||>{\centering\arraybackslash}m{0.25in}|>{\centering\arraybackslash}m{0.3in}|>{\centering\arraybackslash}m{0.25in}|>{\centering\arraybackslash}m{0.25in}|>{\centering\arraybackslash}m{0.25in}|>{\centering\arraybackslash}m{0in} @{}m{0pt}@{}}
			\cline{1-17}
			\multirow{2}{*}{\makecell{Database}} &  \multirow{2}{*}{\makecell{System}} &  \multicolumn{5}{c||}{EEG} & \multicolumn{5}{c||}{EEG$\cdot$EOG} & \multicolumn{5}{c|}{EEG$\cdot$EOG$\cdot$EMG} & \parbox{0pt}{\rule{0pt}{1ex+\baselineskip}} \\ [0ex]  	
			\cline{3-17}
			& & Acc. & $\kappa$ & MF1 & Sens. & Spec. & Acc. & $\kappa$ & MF1 & Sens. & Spec. & Acc. & $\kappa$ & MF1 & Sens. & Spec. & \parbox{0pt}{\rule{0pt}{1ex+\baselineskip}} \\ [0ex]  	
			
			\cline{1-17}
			\multirow{8}{*}{\makecell{~\\SleepEDF-20\\($\pm$ 30 mins)}}   &   \emph{\bf XSleepNet2} &    $\bm{86.3}$ & $\bm{0.813}$ & $\bm{80.6}$ & $\bm{80.2}$ & $\bm{96.4}$  & $\bm{86.4}$ & $\bm{0.813}$ & $\bm{80.9}$ & $\bm{79.9}$ & $\bm{96.2}$   &   & &  & &  & \parbox{0pt}{\rule{0pt}{0ex+\baselineskip}} \\ [0ex]  	
			
			&   \emph{\bf XSleepNet1} &   $\bm{86.0}$ & $\bm{0.810}$ & $\bm{80.0}$ & $\bm{79.6}$ & $\bm{96.3}$ & $85.2$ & $0.798$ & $\bm{79.8}$ & $79.0$ & $95.9$ & $-$ & $-$ & $-$ & $-$ & $-$ & \parbox{0pt}{\rule{0pt}{0ex+\baselineskip}} \\ [0ex]  	
			
			&   \emph{\it Naive Fusion} &  $\it 85.0$ & $\it 0.795$ & $\it 78.8$ & $\it 78.3$ & $\it 96.0$ & $\it 83.4$ & $\it 0.773$ & $\it 77.8$ & $\it 77.1$ & $\it 95.5$ & $-$ & $-$ & $-$ & $-$ & $-$ &  \parbox{0pt}{\rule{0pt}{0ex+\baselineskip}} \\ [0ex]  	
			&   \emph{ARNN+RNN} (SeqSleepNet \cite{Phan2019a}) &   $\it 85.2$ & $\it 0.798$ & $\it 78.4$ & $\it 78.0$ & $\it 96.1$ & $\it 86.0$ & $\it 0.809$ & $\it 79.7$ & $\it 79.2$ & $\it 96.2$ & $-$ & $-$ & $-$ & $-$ & $-$ &  \parbox{0pt}{\rule{0pt}{0ex+\baselineskip}} \\ [0ex]  	
			&   \emph{FCNN+RNN} &   $\it 81.8$ & $\it 0.754$ & $\it 75.6$ & $\it 75.7$ & $\it 95.3$ & $\it 83.5$ & $\it 0.775$ & $\it 77.7$ & $\it 77.2$ & $\it 95.5$ & $-$ & $-$ & $-$ & $-$ & $-$ &  \parbox{0pt}{\rule{0pt}{0ex+\baselineskip}} \\ [0ex]  	
			
			&   DeepSleepNet \cite{Supratak2017} & $-$ & $-$ & $-$  & $-$ & $-$ & $82.0$ & $0.760$ & $76.9$  & $-$ & $-$ & $-$ & $-$ & $-$ & $-$ &$-$ & \parbox{0pt}{\rule{0pt}{0ex+\baselineskip}} \\ [0ex]  	
			&   U-time \cite{Perslev2019} & $-$ & $-$ & $79.0$ & $-$ & $-$ & $-$ & $-$ & $-$ & $-$ & $-$ & $-$ & $-$ & $-$ & $-$ & $-$ & \parbox{0pt}{\rule{0pt}{0ex+\baselineskip}} \\ [0ex]  	
			&  IITNet \cite{Seo2020} & $83.9$ & $0.780$ & $77.6$ & $-$ & $-$ & $-$ & $-$ & $-$ & $-$ & $-$ & $-$ & $-$ & $-$ & $-$ & $-$ & \parbox{0pt}{\rule{0pt}{0ex+\baselineskip}} \\ [0ex]  	
			\cline{1-17}
			
			\multirow{17}{*}{\makecell{~\\SleepEDF-20}}  &   \emph{\bf XSleepNet2} &   $\bm{83.9}$ & $\bm{0.771}$ & $\bm{78.7}$ & $\bm{78.6}$ & $\bm{95.5}$ &  $\bm{83.3}$ & $\bm{0.762}$ & $\bm{77.3}$ & $\bm{76.4}$ & $\bm{95.2}$ &  &   & &  & &  & \parbox{0pt}{\rule{0pt}{0ex+\baselineskip}} \\ [0ex]  	
			
			&   \emph{\bf XSleepNet1} & $\bm{82.2}$ & $\bm{0.750}$ & $\bm{76.4}$ & $\bm{76.8}$ & $\bm{95.2}$ & $\bm{82.3}$ & $\bm{0.747}$ & $\bm{76.1}$ & $\bm{75.2}$ & $\bm{94.9}$ & $-$ & $-$ & $-$ & $-$ & $-$ & \parbox{0pt}{\rule{0pt}{0ex+\baselineskip}} \\ [0ex]  	
			
			&   \emph{\it Naive Fusion} & $\it 80.2$ & $\it 0.723$ & $\it 74.9$ & $\it 75.8$ & $\it 94.7$ & $\it 80.8$ & $\it 0.726$ & $\it 74.7$ & $\it 73.7$ & $\it 94.5$ & $-$ & $-$ & $-$ & $-$ & $-$ & \parbox{0pt}{\rule{0pt}{0ex+\baselineskip}} \\ [0ex]  	
			&   \emph{ARNN+RNN} (SeqSleepNet \cite{Phan2019a}) & $\it 82.2$ & $\it 0.746$ & $\it 74.1$ & $\it 73.9$ & $\it 95.0$& $\it 82.2$ & $\it 0.744$ & $\it 74.2$ & $\it 73.1$ & $\it 94.8$ & $-$ & $-$ & $-$ & $-$ & $-$ & \parbox{0pt}{\rule{0pt}{0ex+\baselineskip}} \\ [0ex]  	
			&   \emph{FCNN+RNN} &  $\it 81.3$ & $\it 0.737$ & $\it 76.0$ & $\it 76.7$ & $\it 95.0$ & $\it 80.1$ & $\it 0.716$ & $\it 73.2$ & $\it 72.6$ & $\it 94.4$ & $-$ & $-$ & $-$ & $-$ & $-$ & \parbox{0pt}{\rule{0pt}{0ex+\baselineskip}} \\ [0ex]  	
			
			& Multitask CNN \cite{Phan2019b} & $81.9$ & $0.740$ & $73.8$ & $-$ & $-$ & $82.3$ & $0.750$ & $74.7$ &  $-$ &  $-$ & $-$  &  $-$ &  $-$ &  $-$ &  $-$ & \parbox{0pt}{\rule{0pt}{0ex+\baselineskip}} \\ [0ex]  	
			&   DeepSleepNet \cite{Phan2019e,Supratak2017} & $80.8$ & $0.731$ & $74.2$  & $-$ & $-$ & $81.9$  & $0.744$ & $75.2$ & $-$ & $-$ & $-$ & $-$ & $-$ & $-$ &$-$ & \parbox{0pt}{\rule{0pt}{0ex+\baselineskip}} \\ [0ex]  	
			& 1-max CNN \cite{phan2018c} & $79.8$ & $0.720$ & $72.0$ &  $-$ &  $-$ &  $-$ &  $-$ &  $-$ &  $-$ &  $-$ &  $-$ &  $-$ &  $-$ &  $-$ & $-$ & \parbox{0pt}{\rule{0pt}{0ex+\baselineskip}} \\ [0ex]  	
			& Attentional RNN \cite{phan2018d} & $79.1$ & $0.700$ & $69.8$ &  $-$ &  $-$ &  $-$ &  $-$ &  $-$ &  $-$ &  $-$ &  $-$ &  $-$ &  $-$ &  $-$ & $-$ & \parbox{0pt}{\rule{0pt}{0ex+\baselineskip}} \\ [0ex]  	
			& Auto-encoder \cite{Tsinalis2016b} & $78.9$ & $-$ & $73.3$ & $-$ & $-$ & $-$ & $-$ & $-$ & $-$ & $-$ & $-$ & $-$ & $-$ & $-$ & $-$ & \parbox{0pt}{\rule{0pt}{0ex+\baselineskip}} \\ [0ex]  	
			&  ResNet \cite{Andreotti2018} &  $-$  &  $0.650$ &  $-$ &  $-$ &  $-$ & $-$ &  $0.680$ &  $-$ &  $-$ &  $-$ &  $-$ &  $-$ &  $-$ &  $-$ & $-$ & \parbox{0pt}{\rule{0pt}{0ex+\baselineskip}} \\ [0ex]  	
			&  VGG-FE \cite{Vilamala2017} & $76.3$ &  $-$ &  $-$ &  $-$ &  $-$ &  $-$ &  $-$ &  $-$ &  $-$ &  $-$ &  $-$ &  $-$ &  $-$ &  $-$ & $-$ & \parbox{0pt}{\rule{0pt}{0ex+\baselineskip}} \\ [0ex]  	
			& CNN \cite{Tsinalis2016} & $74.8$ & $-$ & $69.8$ & $-$ & $-$ & $-$ & $-$ & $-$ & $-$ & $-$ & $-$ & $-$ & $-$ & $-$ & $-$ & \parbox{0pt}{\rule{0pt}{0ex+\baselineskip}} \\ [0ex]  	

			&  {\bf FT-XSleepNet}$^\dagger$ &   $\bm{85.7}$ & $\bm{0.797}$ & $\bm{80.8}$ & $\bm{80.6}$ & $\bm{96.0}$ & $\bm{85.9}$ & $\bm{0.800}$ & $\bm{80.7}$ & $\bm{81.0}$ & $\bm{96.1}$  &   & &  & &  & \parbox{0pt}{\rule{0pt}{0ex+\baselineskip}} \\ [0ex]  	
			
			& FT-SeqSleepNet+$^\dagger$ \cite{Phan2019e} & $85.2$ & $0.789$ & $79.6$ & $-$ & $-$ & $84.3$ & $0.776$ & $77.7$ & $-$ & $-$ & $-$ & $-$ & $-$ & $-$ & $-$ & \parbox{0pt}{\rule{0pt}{0ex+\baselineskip}} \\ [0ex]  	
			& FT-DeepSleepNet+$^\dagger$ \cite{Phan2019e} & $84.4$ & $0.781$ & $78.8$ & $-$ & $-$ & $84.6$ & $0.782$ & $79.0$ & $-$ & $-$ & $-$ & $-$ & $-$ & $-$ & $-$ & \parbox{0pt}{\rule{0pt}{0ex+\baselineskip}} \\ [0ex]  	
			& Person. CNN$^\dagger$ \cite{Mikkelsen2018} & $84.0$ & $-$ & $-$ & $-$ & $-$ & $-$ & $-$ & $-$ & $-$ & $-$ & $-$ & $-$ & $-$ & $-$ & $-$ & \parbox{0pt}{\rule{0pt}{0ex+\baselineskip}} \\ [0ex]  	
			&   VGG-FT$^\dagger$ \cite{Vilamala2017}& $80.3$ &  $-$ &  $-$ &  $-$ &  $-$ & &  $-$ &  $-$ &  $-$ &  $-$ &  $-$ &  $-$ &  $-$ &  $-$ & $-$ & \parbox{0pt}{\rule{0pt}{0ex+\baselineskip}} \\ [0ex]  	
			
			\cline{1-17}
			\multirow{8}{*}{\makecell{~\\SleepEDF-78 \\ ($\pm$ 30 mins)}} &   \emph{\bf XSleepNet2} &  $\bm{84.0}$ & $\bm{0.778}$ & $\bm{77.9}$ & $\bm{77.5}$ & $\bm{95.7}$ &  $\bm{84.0}$ & $\bm{0.778}$ & $\bm{78.7}$ & $\bm{77.6}$ & $\bm{95.7}$  &   & &  & &  & \parbox{0pt}{\rule{0pt}{0ex+\baselineskip}} \\ [0ex]  	
			
			&   \emph{\bf XSleepNet1} &   $\bm{83.6}$ & $\bm{0.773}$ & $\bm{77.8}$ & $\bm{77.7}$ & $\bm{95.7}$ & $\bm{84.0}$ & $\bm{0.777}$ & $\bm{78.4}$ & $77.1$ & $\bm{95.6}$ & $-$ & $-$ & $-$ & $-$ & $-$ & \parbox{0pt}{\rule{0pt}{0ex+\baselineskip}} \\ [0ex]  	
						
			&   \emph{\it Naive Fusion} &   $\it 82.3$ & $\it 0.755$ & $\it 76.2$ & $\it 75.7$ & $\it 95.3$ & $\it 82.5$ & $\it 0.757$ & $\it 76.9$ & $\it 75.8$ & $\it 95.3$ & $-$ & $-$ & $-$ & $-$ & $-$ &  \parbox{0pt}{\rule{0pt}{0ex+\baselineskip}} \\ [0ex]  	
			&   \emph{ARNN+RNN} (SeqSleepNet\cite{Phan2019a}) &   $\it 82.6$ & $\it 0.760$ & $\it 76.4$ & $\it 76.3$ & $\it 95.4$ & $\it 83.8$ & $\it 0.776$ & $\it 78.2$ & $\it 77.4$ & $\it 95.6$ & $-$ & $-$ & $-$ & $-$ & $-$ &  \parbox{0pt}{\rule{0pt}{0ex+\baselineskip}} \\ [0ex]  	
			&   \emph{FCNN+RNN} &   $\it 82.8$ & $\it 0.761$ & $\it 76.6$ & $\it 75.9$ & $\it 95.4$ & $\it 82.7$ & $\it 0.759$ & $\it 76.9$ & $\it 75.5$ & $\it 95.3$ & $-$ & $-$ & $-$ & $-$ & $-$ &  \parbox{0pt}{\rule{0pt}{0ex+\baselineskip}} \\ [0ex]  	
			
			&   U-Time \cite{Perslev2019} &   $-$ & $-$ & $76.0$ & $-$ & $-$ & $-$ & $-$ & $-$ & $-$ & $-$ & $-$ & $-$ & $-$ & $-$ & $-$ &  \parbox{0pt}{\rule{0pt}{0ex+\baselineskip}} \\ [0ex]  	
			&   CNN-LSTM \cite{Perslev2019} &   $-$ & $-$ & $73.0$ & $-$ & $-$ &  $-$ & $-$ & $-$ & $-$ & $-$ & $-$ & $-$ & $-$ & $-$ & $-$ &  \parbox{0pt}{\rule{0pt}{0ex+\baselineskip}} \\ [0ex]  	
			
			& SleepEEGNet \cite{MousaviI2019} &   $80.0$ & $0.730$ & $73.6$ & $-$ & $-$ & $-$ & $-$ & $-$ & $-$ & $-$ & $-$ & $-$ & $-$ & $-$ & $-$ &  \parbox{0pt}{\rule{0pt}{0ex+\baselineskip}} \\ [0ex]  	
			\cline{1-17}
			\multirow{8}{*}{\makecell{SleepEDF-78}}   &   \emph{\bf XSleepNet2} &  $\bm{80.3}$ & $\bm{0.727}$ & $\bm{76.4}$ & $\bm{76.1}$ & $\bm{94.6}$ &  $\bm{80.6}$ & $\bm{0.728}$ & $\bm{76.7}$ & $\bm{75.8}$ & $\bm{94.5}$  &   & &  & &  & \parbox{0pt}{\rule{0pt}{0ex+\baselineskip}} \\ [0ex]  	
			
			&   \emph{\bf XSleepNet1} & $\bm{80.3}$ & $\bm{0.726}$ & $\bm{76.9}$ & $\bm{76.1}$ & $\bm{94.5}$ & $\bm{80.7}$ & $\bm{0.731}$ & $\bm{77.0}$ & $\bm{76.3}$ & $\bm{94.6}$ & $-$ & $-$ & $-$ & $-$ & $-$ & \parbox{0pt}{\rule{0pt}{0ex+\baselineskip}} \\ [0ex]  	
			
			&   \emph{\it Naive Fusion} & $\it 79.1$ & $\it 0.709$ & $\it 75.1$ & $\it 74.3$ & $\it 94.2$ & $\it 79.3$ & $\it 0.711$ & $\it 75.7$ & $\it 74.2$ & $\it 94.1$ & $-$ & $-$ & $-$ & $-$ & $-$ &  \parbox{0pt}{\rule{0pt}{0ex+\baselineskip}} \\ [0ex]  	
			&   \emph{ARNN+RNN} (SeqSleepNet \cite{Phan2019a}) & $\it 79.0$ & $\it 0.708$ & $\it 74.6$ & $\it 74.2$ & $\it 94.2$ & $\it 79.7$ & $\it 0.715$ & $\it 75.7$ & $\it 74.6$ & $\it 94.2$ & $-$ & $-$ & $-$ & $-$ & $-$ &  \parbox{0pt}{\rule{0pt}{0ex+\baselineskip}} \\ [0ex]  	
			&   \emph{FCNN+RNN} & $\it 79.3$ & $\it 0.711$ & $\it 75.1$ & $\it 74.0$ & $\it 94.2$ & $\it 79.8$ & $\it 0.717$ & $\it 76.1$ & $\it 74.9$ & $\it 94.3$ & $-$ & $-$ & $-$ & $-$ & $-$ &  \parbox{0pt}{\rule{0pt}{0ex+\baselineskip}} \\ [0ex]  	
			
			&  Personalized SeqSleepNet$^\dagger$ \cite{Phan2020a} & $79.6$  & $0.706$ & $73.0$ & $71.8$ & $94.2$ & $-$ & $-$ & $-$ & $-$ & $-$ & $-$ & $-$ & $-$ & $-$ & $-$ &  \parbox{0pt}{\rule{0pt}{0ex+\baselineskip}} \\ [0ex]  	
			&  DeepSleepNet \cite{Phan2020a,Supratak2017} &  $78.5$ & $0.702$ & $75.3$ & $75.0$ & $94.1$  & $-$ & $-$ & $-$ & $-$ & $-$ & $-$ & $-$ & $-$ & $-$ & $-$ &  \parbox{0pt}{\rule{0pt}{0ex+\baselineskip}} \\ [0ex]  	
			\cline{1-17}
			\multirow{16}{*}{\makecell{~\\MASS}}  &   \emph{\bf XSleepNet2} &  $\bm{85.2}$ & $\bm{0.788}$ & $\bm{80.6}$ & $\bm{80.2}$ & $\bm{95.8}$ &  $\bm{86.9}$ & $\bm{0.813}$ & $\bm{82.7}$ & $\bm{82.2}$ & $\bm{96.2}$  &   $\bm{87.6}$ & $\bm{0.823}$ & $\bm{83.8}$ & $\bm{83.2}$ & $\bm{96.4}$  & \parbox{0pt}{\rule{0pt}{0ex+\baselineskip}} \\ [0ex]  	
			
			 & \emph{\bf XSleepNet1} &   $\bm{85.1}$ & $\bm{0.788}$ & $\bm{80.6}$ & $\bm{80.4}$ & $\bm{95.8}$ & $\bm{86.8}$ & $\bm{0.812}$ & $\bm{82.6}$ & $\bm{82.1}$ & $\bm{96.2}$ & $\bm{87.5}$ & $\bm{0.821}$ & $\bm{83.7}$ & $\bm{83.1}$ & $\bm{96.4}$ & \parbox{0pt}{\rule{0pt}{0ex+\baselineskip}} \\ [0ex]  	
			
			& \emph{\it Naive Fusion} &   $\it 84.8$ & $\it 0.783$ & $\it 80.2$ & $\it 79.9$ & $\it 95.7$ & $\it 86.2$ & $\it 0.803$ & $\it 81.8$ & $\it 81.2$ & $\it 96.0$ & $\it 86.8$ & $\it 0.812$ & $\it 82.8$ & $\it 82.5$ & $\it 96.2$ & \parbox{0pt}{\rule{0pt}{0ex+\baselineskip}} \\ [0ex]  	
			& \emph{ARNN+RNN} (SeqSleepNet \cite{Phan2019a})&   $\it 84.5$ & $\it 0.778$ & $\it 79.8$ & $\it 79.2$ & $\it 95.6$ & $\it 86.5$ & $\it 0.808$ & $\it 82.4$ & $\it 81.8$ & $\it 96.1$ & $\it 87.0$ & $\it 0.815$ & $\it 83.3$ & $\it 82.7$ & $\it 96.2$ & \parbox{0pt}{\rule{0pt}{0ex+\baselineskip}} \\ [0ex]  	
			& \emph{FCNN+RNN} &   $\it 84.3$ & $\it 0.777$ & $\it 79.5$ & $\it 79.3$ & $\it 95.6$ & $\it 86.0$ & $\it 0.800$ & $\it 81.3$ & $\it 80.8$ & $\it 96.0$ & $\it 86.4$ & $\it 0.806$ & $\it 82.1$ & $\it 81.6$ & $\it 96.1$ & \parbox{0pt}{\rule{0pt}{0ex+\baselineskip}} \\ [0ex]  	
			& DeepSleepNet \cite{Supratak2017,Phan2019a} &  $-$ & $-$  & $-$  & $-$ & $-$  & $-$ & $-$ & $-$ & $-$ & $-$ & $86.4$ & $0.805$ & $82.2$ & $81.8$ & $96.1$ & \parbox{0pt}{\rule{0pt}{0ex+\baselineskip}} \\ [0ex]  	
			
			& Multitask CNN \cite{Phan2019b,Phan2019a} &  $-$&  $-$&  $-$&  $-$&  $-$& $-$ & $-$ & $-$ & $-$ & $-$ & $83.6$ & $0.766$ & $77.9$ & $77.4$ & $95.3$ & \parbox{0pt}{\rule{0pt}{0ex+\baselineskip}} \\ [0ex]  	
			& Attentional RNN \cite{Phan2019a,phan2018d}  & $-$  & $-$  & $-$  & $-$  & $-$  & $-$ & $-$ & $-$ & $-$ & $-$ & $83.6$ & $0.766$ & $78.4 $ & $78.0$ & $95.3$ & \parbox{0pt}{\rule{0pt}{0ex+\baselineskip}} \\ [0ex]  	
			& 1-max CNN \cite{Phan2019a,phan2018c}  &  $-$ &  $-$ &  $-$ &  $-$ &  $-$ & $-$ & $-$ & $-$ & $-$ & $-$ & $82.7$ & $0.754$ & $77.6 $ & $77.8$ & $95.1$ & \parbox{0pt}{\rule{0pt}{0ex+\baselineskip}} \\ [0ex]  	
			
			& CNN \cite{Chambon2018, Phan2019a}  &  $-$ &  $-$ &  $-$ &  $-$ &  $-$ & $-$ & $-$ & $-$ & $-$ & $-$ & $79.9$ & $0.726$ & $76.7$ & $80.0$ & $ 95.0$ &  \parbox{0pt}{\rule{0pt}{0ex+\baselineskip}} \\ [0ex]  	
			& CNN \cite{Tsinalis2016,Phan2019a}  & $-$  & $-$  & $-$  & $-$  & $-$  & $-$ & $-$ & $-$ & $-$ & $-$ & $77.9$ & $0.680$ & $70.4$ & $69.4$ & $ 93.5$ & \parbox{0pt}{\rule{0pt}{0ex+\baselineskip}} \\ [0ex]  	
			& ResNet \cite{Andreotti2018}  & $-$ & $0.670$ &  $-$ &  $-$ &  $-$ & $-$ & $0.720$ & $-$ & $-$ & $-$ & $-$ & $0.740$ & $-$ & $-$ & $-$ & \parbox{0pt}{\rule{0pt}{0ex+\baselineskip}} \\ [0ex]  	
			
			&  IITNet$^\dagger$ \cite{Seo2020} & $86.3$ & $0.790$ & $80.5$ & $-$ & $-$ & $-$ & $-$ & $-$ & $-$ & $-$ & $-$ & $-$ & $-$ & $-$ & $-$ & \parbox{0pt}{\rule{0pt}{0ex+\baselineskip}} \\ [0ex]  	
			& DeepSleepNet$^\dagger$ \cite{Supratak2017} &  $-$ & $-$  & $-$  & $-$ &  $-$ & $86.2$ & $0.800$ & $81.7$ & $-$ & $-$ & $-$ & $-$ & $-$ & $-$ & $-$ &\parbox{0pt}{\rule{0pt}{0ex+\baselineskip}} \\ [0ex]  	
			& DNN+RNN$^\dagger$ \cite{Dong2017} & $-$  & $-$  & $-$  & $-$  & $-$  & $85.9$ & $-$ & $80.5$ & $-$ & $-$ & $-$ & $-$ & $-$ & $-$ & $-$ & \parbox{0pt}{\rule{0pt}{0ex+\baselineskip}} \\ [0ex]  	
			& DNN$^\dagger$ \cite{Dong2017} &  $-$ & $-$  & $-$  & $-$  & $-$  & $81.6$ & $-$ & $77.2$ & $-$ & $-$ & $-$ & $-$ & $-$ & $-$ & $-$ & \parbox{0pt}{\rule{0pt}{0ex+\baselineskip}} \\ [0ex]  	
			\cline{1-17}
			\multirow{7}{*}{\makecell{~\\Physio2018}}  &   \emph{\bf XSleepNet2} &  $\bm{80.3}$ & $\bm{0.732}$ & $\bm{78.6}$ & $\bm{78.7}$ & $\bm{94.6}$ &  $\bm{81.4}$ & $\bm{0.746}$ & $\bm{79.9}$ & $\bm{79.9}$ & $\bm{94.9}$  &   $\bm{81.1}$ & $\bm{0.742}$ & $\bm{79.4}$ & $\bm{79.5}$ & $\bm{94.8}$ & \parbox{0pt}{\rule{0pt}{0ex+\baselineskip}} \\ [0ex]  	
			
			&   \emph{\bf XSleepNet1} &   $\bm{80.3}$ & $\bm{0.731}$ & $\bm{78.5}$ & $\bm{78.4}$ & $\bm{94.6}$ & $\bm{81.2}$ & $\bm{0.744}$ & $\bm{79.6}$ & $\bm{79.7}$ & $\bm{94.8}$ &  $\bm{81.1}$ & $\bm{0.742}$ & $\bm{79.5}$ & $\bm{79.3}$ & $\bm{94.8}$ & \parbox{0pt}{\rule{0pt}{0ex+\baselineskip}} \\ [0ex]  	
						
			&   \emph{\it Naive Fusion} &  $\it 80.0$ & $\it 0.727$ & $\it 78.2$ & $\it 78.1$ & $\it 94.5$ & $\it 80.7$ & $\it 0.736$ & $\it 79.0$ & $\it 78.9$ & $\it 94.7$ & $\it 80.7$ & $\it 0.737$ & $\it 79.2$ & $\it 79.0$ & $\it 94.7$ & \parbox{0pt}{\rule{0pt}{0ex+\baselineskip}} \\ [0ex]  	
			&   \emph{ARNN+RNN} (SeqSleepNet \cite{Phan2019a})&  $\it 79.4$ & $\it 0.719$ & $\it 77.6$ & $\it 77.5$ & $\it 94.3$ & $\it 80.5$ & $\it 0.734$ & $\it 78.9$ & $\it 78.9$ & $\it 94.6$  & $\it 80.4$ & $\it 0.733$ & $\it 78.8$ & $\it 78.8$ & $\it 94.6$ & \parbox{0pt}{\rule{0pt}{0ex+\baselineskip}} \\ [0ex]  	
			&   \emph{FCNN+RNN} &   $\it 79.7$ & $\it 0.723$ & $\it 77.8$ & $\it 77.5$ & $\it 94.4$ & $\it 81.0$ & $\it 0.741$ & $\it 79.2$ & $\it 79.1$ & $\it 94.8$ & $\it 80.7$ & $\it 0.738$ & $\it 79.2$ & $\it 79.2$ & $\it 94.7$ & \parbox{0pt}{\rule{0pt}{0ex+\baselineskip}} \\ [0ex]  	
			
			&   U-Time \cite{Perslev2019} &   $-$ & $-$ & $77.0$ & $-$ & $-$ & $-$ & $-$ & $-$ & $-$ & $-$ &  $-$ & $-$ & $77.0$ & $-$ & $-$ & \parbox{0pt}{\rule{0pt}{0ex+\baselineskip}} \\ [0ex]  	
			&   CNN-LSTM \cite{Perslev2019} &   $-$ & $-$ & $77.0$ & $-$ & $-$ & $-$ & $-$ & $-$ & $-$ & $-$ &  $-$ & $-$ & $-$ & $-$ & $-$  & \parbox{0pt}{\rule{0pt}{0ex+\baselineskip}} \\ [0ex]  	
			\cline{1-17}
			\multirow{6}{*}{\makecell{~\\SHHS}}  &   \emph{\bf XSleepNet2} &  $\bm{87.6}$ & $\bm{0.826}$ & $\bm{80.7}$ & $79.7$ & $\bm{96.5}$ &  $\bm{88.8}$ & $\bm{0.843}$ & $\bm{81.8}$ & $80.8$ & $\bm{96.8}$  &  $\bm{89.1}$ & $\bm{0.847}$ & $\bm{82.3}$ & $\bm{81.2}$ & $\bm{96.9}$   & \parbox{0pt}{\rule{0pt}{0ex+\baselineskip}} \\ [0ex]  	
			
			 & \emph{\bf XSleepNet1} &   $\bm{87.5}$ & $\bm{0.826}$ & $\bm{81.0}$ & $\bm{80.4}$ & $\bm{96.5}$ & $\bm{88.8}$ & $\bm{0.843}$ & $\bm{82.0}$ & $81.3$ & $\bm{96.8}$ & $\bm{89.1}$ & $\bm{0.847}$ & $\bm{82.2}$ & $\bm{81.4}$ & $\bm{96.9}$ & \parbox{0pt}{\rule{0pt}{0ex+\baselineskip}} \\ [0ex]  	
						
			& \emph{Naive Fusion} &   $\it 87.5$ & $\it 0.825$ & $\it 80.7$ & $\it 79.8$ & $\it 96.5$ & $\it 88.4$ & $\it 0.839$ & $\it 81.7$ & $\it 81.6$ & $\it 96.8$ & $\it 88.8$ & $\it 0.843$ & $\it 81.7$ & $\it 80.8$ & $\it 96.8$ & \parbox{0pt}{\rule{0pt}{0ex+\baselineskip}} \\ [0ex]  	
			& \emph{ARNN+RNN} (SeqSleepNet \cite{Phan2019a})&   $\it 86.5$ & $\it 0.811$ & $\it 78.5$ & $\it 76.9$ & $\it 96.1$ & $\it 88.4$ & $\it 0.837$ & $\it 80.7$ & $\it 79.6$ & $\it 96.7$ & $\it 88.4$ & $\it 0.838$ & $\it 80.1$ & $\it 78.5$ & $\it 96.7$ & \parbox{0pt}{\rule{0pt}{0ex+\baselineskip}} \\ [0ex]  	
			& \emph{FCNN+RNN} &   $\it 86.7$ & $\it 0.813$ & $\it 79.5$ & $\it 78.1$ & $\it 96.2$ & $\it 88.0$ & $\it 0.831$ & $\it 80.5$ & $\it 79.2$ & $\it 96.6$ & $\it 88.1$ & $\it 0.832$ & $\it 80.9$ & $\it 79.7$ & $\it 96.6$ &  \parbox{0pt}{\rule{0pt}{0ex+\baselineskip}} \\ [0ex]  	
			
			& CNN \cite{Sors2018} &   $86.8$ & $0.810$ & $78.5$ & $-$ & $95.0$ & $-$ & $-$ & $-$ & $-$ & $-$ & $-$ & $-$ & $-$ & $-$ & $-$ & \parbox{0pt}{\rule{0pt}{0ex+\baselineskip}} \\ [0ex]  	
			&  IITNet \cite{Seo2020} & $86.7$ & $0.810$ & $79.8$ & $-$ & $-$ & $-$ & $-$ & $-$ & $-$ & $-$ & $-$ & $-$ & $-$ & $-$ & $-$ & \parbox{0pt}{\rule{0pt}{0ex+\baselineskip}} \\ [0ex]  	
			\cline{1-17}
		\end{tabular}
	\end{center}
	\label{tab:performance}
\end{table*}

\vspace{-0.1cm}
\subsection{Experimental results}
\label{sec:experiment_results}

To give an overview of the performances of XSleepNet1, XSleepNet2, and the baselines, we collate and contrast their overall performances across all the experimental databases across different channel combinations in Fig. \ref{fig:overall_performance}. 

Fig. \ref{fig:overall_performance} reveals several compelling patterns. First, between the two single-view baselines, \emph{ARNN+RNN} and \emph{FCNN+RNN}, the former often results in better performance when the data size is relatively small, for example in SleepEDF-20 and MASS. The opposite is commonly seen with larger databases, such as in Physio2018 and SHHS (EEG). These patterns can be partly explained by the difference in their model footprints: the \emph{ARNN+RNN} with single-channel EEG input has about $1.6\!\times\!10^5$ parameters in total, 35 times fewer than $5.6\!\times\!10^6$ parameters in the \emph{FCNN+RNN}. As a result, \emph{FCNN+RNN} is prone to overfitting on the smaller databases (e.g. SleepEDF-20) whereas \emph{ARNN+RNN} is less capable of dealing with the larger databases (e.g. Physio2018). However, it should be stressed that model size is not the only explanation. Indeed, the patterns in SHHS (EEG$\cdot$EOG and EEG$\cdot$EOG$\cdot$EMG) seem to be counter-intuitive. As indicated in Section \ref{ssec:architecture}, the CNN-based and RNN-based models tend to capture different kinds of patterns from their input; therefore, performance discrepancies are expected on different sleep stages \cite{Phan2019a,Phan2019e}. This suggests that how well an individual model performs also depends on the sleep structure of a target cohort and/or the channel combination used. 

Second, simple concatenation of the two views in \emph{Naive Fusion} leads to different, potentially diverging results. One can observe clear performance improvements over the two single-view baselines in some cases, such as MASS (EEG) and Physio2018 (EEG). In other cases, such as MASS (EEG$\cdot$EOG) and Physio2018 (EEG$\cdot$EOG), \emph{Naive Fusion}'s performance appears to be averaged between the two single-view baselines. There are a few extreme cases where \emph{Naive Fusion} is inferior to both single-view baselines, such as in SleepEDF-20 (EEG) and SleepEDF-78 (EEG$\cdot$EOG). These diverging patterns indicate that a naive fusion strategy cannot guarantee performance gain, potentially owing to the asynchronous learning behavior of the two views. 

Third, via the generalization-/overfitting-aware learning scheme, the XSleepNets can coordinate the learning pace of the views, consolidate the representation power of the views in the joint representation, and convert them into performance gains. Apart from XSleepNet1's modest performance on the smallest database SleepEDF-20 (which will be further discussed), improvements over both the single-view and the naive fusion baselines are consistently seen across all the experimental databases and channel combinations. 

In Table \ref{tab:performance}, we lay out the detailed performance of XSleepNet1, XSleepNet2, the developed baselines, and existing works that used the experimental databases. Particularly, for SleepEDF-20 and SleepEDF-78, we experimented with two common ways these databases were used in literature: (1) only \emph{in-bed} parts of the recordings were used as recommended in \cite{Imtiaz2014,Imtiaz2015}; (2) 30 minutes of data before and after in-bed parts were further included in the experiments following the initiation in \cite{Supratak2017}. 

There are a few important points. First, the results in Table \ref{tab:performance} indicate that the single-view baselines developed in this work are competent models for sleep staging. The rationale is that they all adhere to the state-of-the-art sequence-to-sequence sleep staging framework \cite{Phan2019a,Phan2019e}. Under a similar experimental condition, \emph{ARNN+RNN} outperforms all the existing works across all the scenarios, except U-time \cite{Perslev2019} on SleepEDF-20. Although the other single-view baseline, \emph{FCNN+RNN}, underperforms on the smallest database, SleepEDF-20, likely owing to its large model footprint, its performance improves when the data size increases. On all the databases but SleepEDF-20, its performance are often better than, and occasionally comparable to, those of the prior works. Second, the results in Table \ref{tab:performance} confirm the efficacy of XSleepNet1 and XSleepNet2. 
They outperform not only the baselines but also prior works across all the databases and channel combinations. 
On average, XSleepNet1 improves the overall accuracy by $0.9\%$, $0.6\%$, and $1.1\%$ absolute over the \emph{Naive Fusion}, \emph{ARNN+RNN}, and \emph{FCNN+RNN} baselines, respectively. The corresponding accuracy gains are even higher with XSleepNet2, reaching $1.2\%$, $0.9\%$, and $1.4\%$, respectively. Although we do not compute such a performance gain over previous works covered in Table \ref{tab:performance} due to incomplete reported results, improvements with large margins can be expected.  To demonstrate that XSleepNet also performs well in sleep transfer learning tasks, we carried out the transfer learning tasks as proposed in \cite{Phan2019e}. More specifically, we pretrained XSleepNet on the MASS database (i.e., the source domain data) (190 subjects used for training and 10 subjects used for validation) and finetuned the pretrained network on the SleepEDF-20 database (i.e., the target domain). Of note, XSleepNet2's approximation approach was used in pretraining whereas XSleepNet1's approximation approach was used in finetuning. This was because we observed that finetuning could converge quickly after a few iterations, XSleepNet2's approximation was inappropriate due to its warmup period. This system is denoted as FT-XSleepNet in Table \ref{tab:performance}. As can be seen, FT-XSleepNet outperforms the two transfer-learning approaches proposed in \cite{Phan2019e}, FT-SeqSleepNet+ and FT-DeepSleepNet+, improving the overall accuracy by $0.5\%$ and $1.3\%$ on SleepEDF-20 (EEG) and by $1.6\%$ and $1.3\%$ on SleepEDF-20 (EEG$\cdot$EOG), respectively.

Inspection on the class-wise performances also reveals that XSleepNet1 and XSleepNet2 lead improvements over the baselines in many databases and channel combinations. More importantly, oftentimes these improvements are spread over the five sleep stages rather than biasing towards some particular ones (see Appendix C).

\vspace{-0.35cm}
\section{Discussion}
\label{ssec:discussion}

\begin{figure} [!t]
	\centering
	\includegraphics[width=0.9\linewidth]{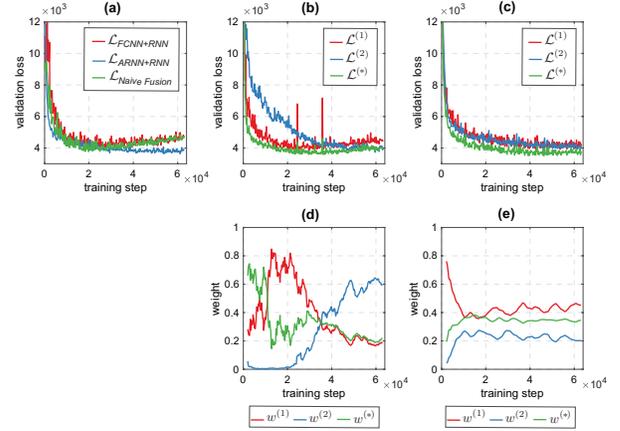}
	\vspace{-0.25cm}
	\caption{Progression of the validation losses and the adaptive loss weights during the training course of one MASS cross-validation fold. (a) The validation losses of the \emph{FCNN+RNN}, \emph{ARNN+RNN}, and \emph{Naive Fusion} baselines; (b) The validation losses of the three classification branches of XSleepNet1 and (d) their respective adaptive loss weights; (c) The validation losses of the three classification branches of XSleepNet2 and (e) their respective adaptive loss weights. It should be emphasized that the adaptive loss weights in (c) were denoised with a 10-point moving average filter before plotting; the original ones are expectedly much noisier.}
	\label{fig:loss_vs_weight}
	\vspace{-0.25cm}
\end{figure}

As mentioned in Section \ref{ssec:gradient_bending}, XSleepNet2 is devised to overcome the unstable learning behavior of XSleepNet1, especially when the data size is small. Indeed, the stability of the two approaches can be illuminated via their performances shown in Table \ref{tab:performance}. Both XSleepNet1 and XSleepNet2 work remarkably well when the data size is medium or large (i.e., all the databases but SleepEDF-20) and the latter's performance appears to be better than or equivalent to that of the former in most cases, except for MASS (EEG$\cdot$EOG$\cdot$EMG). 
Their discrepancy becomes clear when inspecting their performance on the smaller database, SleepEDF-20. In this case, although XSleepNet1 still maintains its performance advantages over the \emph{Naive Fusion} baseline, it does not necessarily outperform the best small-footprint baseline \emph{ARNN+RNN}, particularly when its overall accuracy falls short of the baseline's with a gap of $0.8\%$ on SleepEDF-20 (EEG$\cdot$EOG). In contrast, XSleepNet2's superiority is ubiquitous, not only over all the baselines but also over the XSleepNet1. XSleepNet1's instability is most likely due to the noisy loss weights (cf. Fig. \ref{fig:loss_vs_weight}) produced by its first-order approach used in generalization/overfitting measure approximation. One possibility to overcome this instability is to utilize the multiple outputs of the network in a self-ensemble fashion (see Appendix D).

To see how the loss weights were adapted for gradient blending during the training course of the XSleepNets, we take one cross-validation fold of the MASS database as an example (illustrated in Fig. \ref{fig:loss_vs_weight}). In case of the single-view training with \emph{ARNN+RNN} and \emph{FCNN+RNN}, even though the former's validation loss did not show signs of overfitting, the latter's with a large model footprint converged and started overfitting after $10^5$ training steps (cf. Fig. \ref{fig:loss_vs_weight} (a)). As a consequence, when the two views were trained jointly with a simple fusion strategy in \emph{Naive Fusion}, the overfitting network stream kept learning at the same rate, impairing the joint-view representation and causing overfitting (even though less serious) in the joint classification branch (it can be roughly thought as being the average of the two views). In contrast, during the training of XSleepNet1 (cf. Fig. \ref{fig:loss_vs_weight} (b) and (d)), the view in red that was converging faster was initially associated with an increasingly large weight while a small weight was assigned to the one in blue (the slower one). Note that the view in blue was still able to learn due to the gradient flow coming from the joint classification branch (the green curve). The turning point was when the view in red converged and started overfitting shortly afterwards; its weights descended, impeding the learning and preventing it from overfitting the data at the regular pace. The decreasing weight of the view in red was gradually transferred to and accelerate the learning of the view in blue. As a result, the joint learning process yields the joint representation which is more robust to overfitting and more generalizable than that learned by the \emph{Naive Fusion} baseline. On the other hand, XSleepNet2's losses and weights (cf. Fig. \ref{fig:loss_vs_weight} (c) and (e)) reveal different learning behavior from that of XSleepNet1. Apart from being much smoother, the loss weights appear to be adapted to synchronize the learning pace of the two views and then maintain this synchronization throughout rather than behaving in an turn-taking fashion as in XSleepNet1.

Out of five experimental databases, only on SHHS, the largest database, did we see similar behavior between the \emph{Naive Fusion} and the XSleepNets. In other words, the \emph{Naive Fusion} model resulted in consistently better results than the single-view baselines (cf. Fig. \ref{fig:overall_performance} and Table \ref{tab:performance}). The reason is when the data becomes large enough, it imposes a strong regularization on the network. As a consequence, the networks, either in single-view training or in joint training, did not experience overfitting as with smaller databases, evidenced by the validation losses in Fig. \ref{fig:loss_vs_weight}. We anticipate that the model size can be safely increased in this case.

\begin{figure} [!t]
	\centering
	\includegraphics[width=0.9\linewidth]{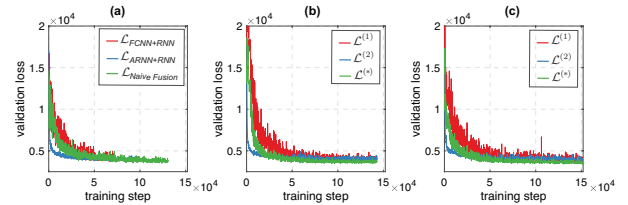}
	\vspace{-0.25cm}
	\caption{The validation losses during the training course of the SHHS database. (a) The validation losses of the \emph{FCNN+RNN}, \emph{ARNN+RNN}, and \emph{Naive Fusion} baselines; (b) The validation losses of the three classification branches of XSleepNet1; (c) The validation losses of the three classification branches of XSleepNet2.}
	\label{fig:validation_loss_shhs}
	\vspace{-0.25cm}
\end{figure}

It is worth mentioning that even though XSleepNet has two network streams, its model footprint ($5.8\times 10^6$ parameters) is still 4 times smaller than it of the popular DeepSleepNet ($22.9\times10^6$ parameters). In addition, utilizing multiple EEG channels, when available, of a sleep database would further improve XSleepNet's \cite{Chambon2018}. An interesting question is whether to combine the EEG channels as in \cite{Chambon2018} or consider them as different views and using our proposed muliti-view approach for channel combination. Although we experimented the XSleepNets on automatic sleep staging with PSG signals, the method is generic enough to serve sleep analysis with other modalities, especially when multimodal data are available \cite{Zhai2020, Olesen2019}. It is also applicable to other applications where the target signals are inherently multi-view. One example is audio/speech in which raw audio signals \cite{Palaz2015} can be combined with its derived representations, such as mel-scale spectrogram \cite{Amodei2016} and gammatone spectrogram \cite{Schlueter2007}, for recognition tasks. Another example is computer vision in which different image channels, such as luminance, chrominance, depth, and optical flow, are essentially multi-view data \cite{Tian2019, Chen2017}.

\vspace{-0.25cm}
\section{Conclusions}
\label{sec:conclusion}

In this paper, we presented XSleepNet, a sequence-to-sequence network architecture for automatic sleep staging, that is capable of learning from both raw signal and time-frequency input at the same time. The network architecture accommodates two network streams, one for each input view. XSleepNet is principally designed to be robust to training data size, complementary between the constituent network streams, and aware of generalization and overfitting behavior of its network streams. The network can be trained in such a way that learning on the generalizing network stream is encouraged while that on the overfitting one is discouraged. Two approaches were introduced to approximate generalization/overfitting measure on classification branches of the network and produce the respective weights for gradient blending, resulting in two models, XSleepNet1 and XSleepNet2. By regulating the weights, and hence the learning pace of the network streams, XSleepNet1 and XSleepNet2 yielded joint features which represent the underlying data distribution better than those learned by the single-view baselines as well as the multi-view baseline following a naive fusion approach. Empirical evaluation showed that they not only delivered more favorable results than the baselines but also outperformed existing work on five databases of different sizes. Between the two models, XSleepNet2, relying on a second-order approximation algorithm, resulted in more stable performance than XSleepNet1, using a first-order approximation algorithm.

	\vspace{-0.25cm}
	\ifCLASSOPTIONcompsoc
	\section*{Acknowledgments}
	\else
	\section*{Acknowledgment}
	\fi
	
	The authors would like to acknowledge and thank Dr. Long Tran-Thanh at the University of Warwick, UK for his helpful discussion and comments. This research received funding from the Flemish Government (AI Research Program). Maarten De Vos is affiliated to Leuven.AI - KU Leuven institute for AI, B-3000, Leuven, Belgium. The study was approved by Clinical Trials and Research Governance, Churchill Hospital - Oxford University Hospitals, Oxford, UK. Data were provided by the Center for Sleep and Wake Disorders at MCH Westeinde Hospital, Den Haag, The Netherlands; the Center for Advanced Research in Sleep Medicine, Montreal, Canada; the Sleep Lab, the Computational Clinical Neurophysiology Laboratory, and the Clinical Data Animation Center at Massachusetts General Hospital, MA, USA; and the Division of Sleep and Circadian Disorders at Brigham and Women's Hospital, MA, USA.
	\ifCLASSOPTIONcaptionsoff
	\newpage
	\fi

	
	
	%
	\vspace{-0.25cm}
	\bibliographystyle{IEEEbib}
	\bibliography{bibliography}
	\vspace{-1cm}
	\vskip 0pt plus -1fil
	\begin{IEEEbiography}[{\includegraphics[width=1in,height=1.25in,clip,keepaspectratio]{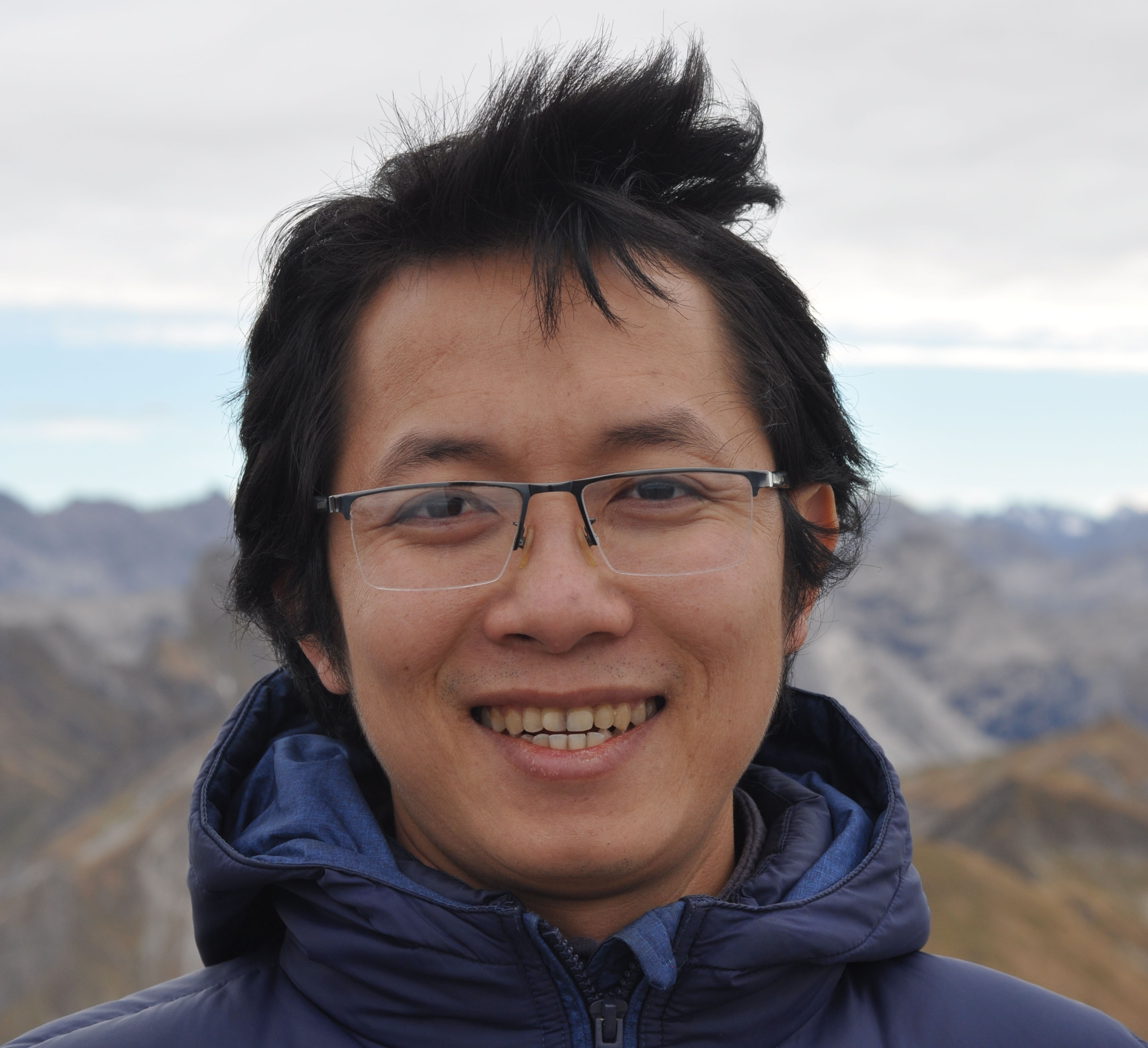}}]{Huy Phan}
		received the M.Eng. degree from Nanyang Technological University, Singapore, in 2012, and the Dr.-Ing. degree in computer science from University of L\"ubeck, Germany, in 2017. From 2017 to 2018, he was a Postdoctoral Research Assistant with University of Oxford, UK. From 2019 to 2020, he was a Lecturer at University of Kent, UK. In April 2020, he joined Queen Mary University of London, UK, where he is a Lecturer in Artificial Intelligence. His research interests include machine learning and signal processing with a special focus on audio and biosignal analysis. In 2018, he received the Bernd Fischer Award for the best PhD thesis from University of L\"ubeck. In 2021, he was awarded Benelux's IEEE-EMBS Best Paper Award 2019-20. 
	\end{IEEEbiography}
	\vskip 0pt plus -1fil
		\begin{IEEEbiography}[{\includegraphics[width=1in,height=1.25in,clip,keepaspectratio]{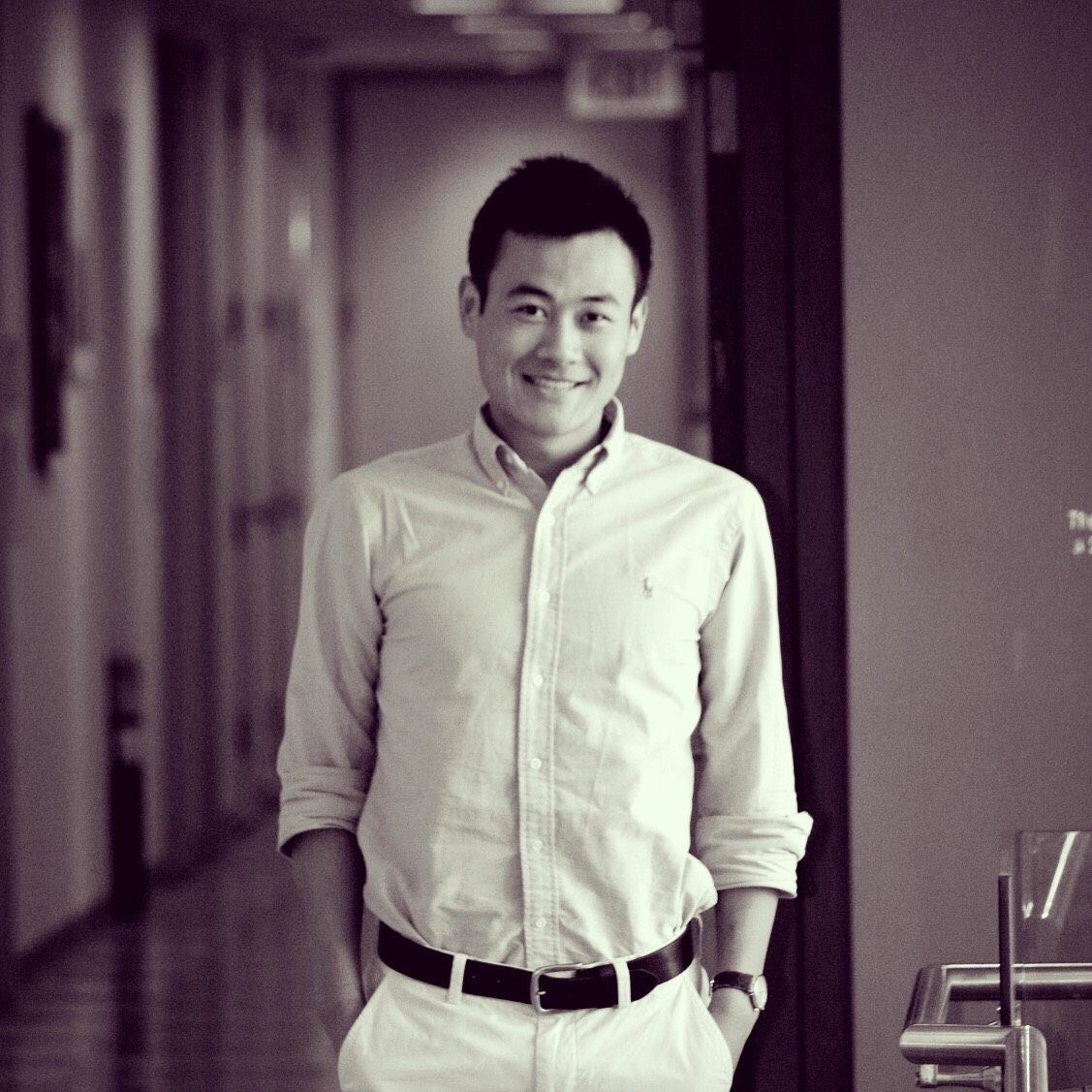}}]{Oliver Y. Ch\'en}
		reads biomedical engineering in the Institute of Biomedical Engineering at University of Oxford and is an Honorary Research Fellow in the Division of Biosciences at University College London. He received his Master’s in Biostatistics from Johns Hopkins University, Master's in Mathematical Statistics from Washington State University, and Ph.D. in Engineering Science at University of Oxford. He was a Research Fellow in Neuroscience and Psychology at Yale University; he gave the Si-Shui Lecture and received the Louise I. and Thomas D. Dublin Award for the Advancement of Epidemiology and Biostatistics. 
	\end{IEEEbiography}
	\vspace{-1cm}
	\vskip 0pt plus -1fil
	\begin{IEEEbiography}[{\includegraphics[width=1in,height=1.25in,clip,keepaspectratio]{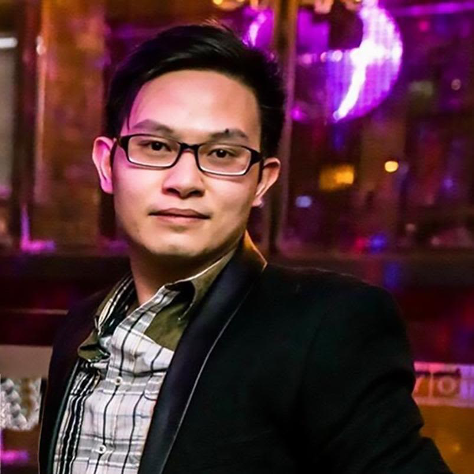}}]{Minh C. Tran}
		received the B.Eng. (Hons) degree in Bioengineering from the University of Sheffield, United Kingdom in 2017. He is currently a graduate reading for D.Phil. in Engineering Science at the University of Oxford. He is a member of the Nuffield Division of Anaesthetics, University of Oxford. His research interests include biosignal modelling and processing, respiratory and intensive care medical devices, and entrepreneurship.
	\end{IEEEbiography}
	\vspace{-1cm}
	\vskip 0pt plus -1fil
	\begin{IEEEbiography}[{\includegraphics[width=1in,height=1.25in,clip,keepaspectratio]{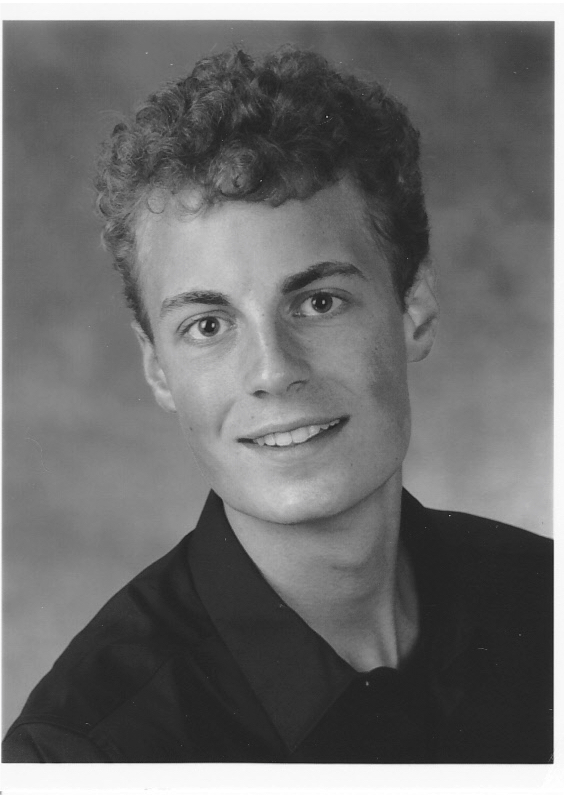}}]{Philipp Koch}
		received the B.Sc. and M.Sc. degrees in medical engineering science in 2013 and 2015, respectively, from the University of L\"ubeck, L\"ubeck, Germany, where he is currently working toward the Ph.D. degree. He is a Research Associate at the Institute for Signal Processing, University of L\"ubeck. His research interests include machine learning, biosignal analysis, audio/acoustic signal processing, and sparse MRI reconstruction.
	\end{IEEEbiography}
	\vspace{-1cm}
	\vskip 0pt plus -1fil
	\begin{IEEEbiography}[{\includegraphics[width=1in,height=1.25in,clip,keepaspectratio]{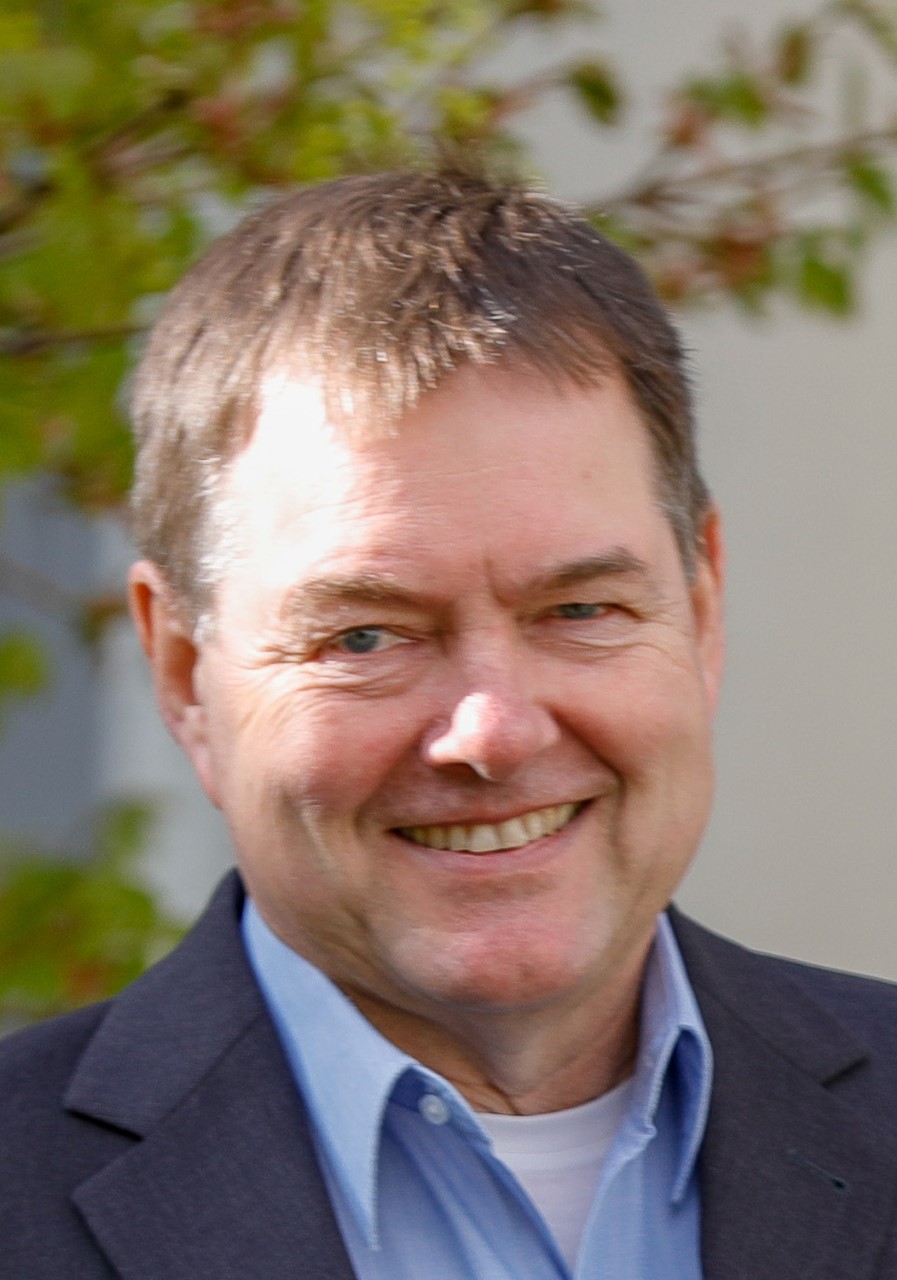}}]{Alfred Mertins}
		received the Dipl.-Ing. degree from the University of Paderborn, Paderborn, Germany, in 1984, the Dr.-Ing. degree in electrical engineering and the Dr.-Ing. Habil. degree in telecommunications from Hamburg University of Technology, Hamburg, Germany, in 1991 and 1994, respectively. From 1986 to 1991, he was a Research Assistant with Hamburg University of Technology, and from 1991 to 1995, he was a Senior Scientist with the Microelectronics Applications Center Hamburg, Germany. From 1996 to 1997, he was with the University of Kiel, Germany, and from 1997 to 1998, with the University of Western Australia, WA, Australia. In 1998, he joined the University of Wollongong, Wollongong, NSW, Australia, where he was an Associate Professor of electrical engineering. From 2003 to 2006, he was a Professor with the Faculty of Mathematics and Science, University of Oldenburg, Oldenburg, Germany. In November 2006, he joined the University of L\"ubeck, L\"ubeck, Germany, where he is a Professor and the Director of the Institute for Signal Processing. His research interests include speech, audio, and image processing; wavelets and filter banks; pattern recognition; and digital communications. 
	\end{IEEEbiography}
	\vspace{-1cm}
	\vskip 0pt plus -1fil
	\begin{IEEEbiography}[{\includegraphics[width=1in,height=1.25in,clip,keepaspectratio]{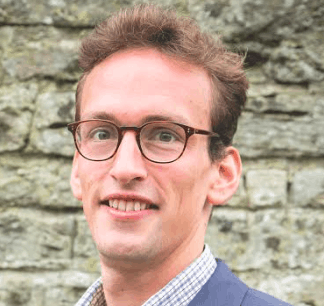}}]{Maarten De Vos}
		has a joint appointment as Associate Professor in the Departments of Engineering and Medicine at KU Leuven after being Associate Professor at the University of Oxford, United Kingdom, and Junior Professor at the University of Oldenburg, Germany. He obtained an MSc (2005) and PhD (2009) in Electrical Engineering from KU Leuven, Belgium. His academic work focuses on AI for health, innovative biomedical monitoring and signal analysis for daily life applications, in particular the derivation of personalised biosignatures of patient health from data acquired via wearable sensors and the incorporation of smart analytics into unobtrusive systems. His pioneering research in the field of mobile real-life brain-monitoring has won several innovation prices, among which the prestigious Mobile Brain Body monitoring prize in 2017. In 2019, he was awarded the Martin Black Prize for the best paper in Physiological Measurement.
	\end{IEEEbiography}
	%
	
	%
	%
	%
	
	
	

\end{document}